\documentclass[reqno,twocolumn]{article}

\usepackage[paperwidth=209.6mm, paperheight=279.4mm, left=15mm, right=15mm, top=30mm, bottom=30mm,asymmetric]{geometry}

\usepackage{amsfonts,amssymb,amsmath,amscd,amsthm,mathtools}

\usepackage{lmodern}
\usepackage[bold]{sourcesanspro}
\usepackage{crimson}
\usepackage[dvipsnames]{xcolor}
\usepackage{authblk}
\usepackage{sectsty}
\usepackage{titlesec}
\usepackage[singlespacing]{setspace}
\usepackage[font={footnotesize,sf},labelfont={bf},labelsep={period}]{caption}
\usepackage[utf8]{inputenc}

\usepackage{enumitem}

\usepackage{natbib} 

\usepackage{hyperref}
\usepackage{empheq}

\definecolor{redheadings}{HTML}{DF3E31}
\definecolor{blackheadings}{HTML}{000000}

\allsectionsfont{\sffamily\normalsize\color{blackheadings}}
\subsectionfont{\normalsize\slshape\sffamily\bfseries\small\color{blackheadings}}
\subsubsectionfont{\normalsize\slshape\sffamily\mdseries\small\color{blackheadings}}

\titlespacing\section{0pt}{12pt plus 4pt minus 2pt}{0pt plus 2pt minus 2pt}
\titlespacing\subsection{0pt}{12pt plus 4pt minus 2pt}{0pt plus 2pt minus 2pt}
\titlespacing\subsubsection{0pt}{12pt plus 4pt minus 2pt}{0pt plus 2pt minus 2pt}

\titlelabel{\thetitle.\hspace{4pt}}
\title{\sffamily The Two Kinds of Free Energy and the Bayesian Revolution}

\author[1,2]{\sffamily Sebastian Gottwald}
\author[1]{\sffamily Daniel A. Braun}

\affil[1]{Institute of Neural Information Processing, Ulm University, 89081 Ulm, Germany}
\affil[2]{sebastian.gottwald@uni-ulm.de}
\date{}           
\newcommand{\argmax}{\mathop{\mathrm{argmax}}}
\newcommand{\argmin}{\mathop{\mathrm{argmin}}}
\newcommand{\DKL}{D_\mathrm{KL}}

\newcommand{\eqq}{\hspace{1pt}{=}\hspace{2pt}}
\newcommand{\eeqq}{\hspace{2pt}{=}\hspace{2pt}}

\begin{document}

\twocolumn[
\begin{@twocolumnfalse}
    \maketitle

\begin{quote}
\section*{Abstract}
    The concept of free energy has its origins in 19th century thermodynamics, but has recently found its way into the behavioral and neural sciences, where it has been promoted for its wide applicability and has even been suggested as a fundamental principle of understanding intelligent behavior and brain function. We argue that there are essentially two different notions of free energy in current models of intelligent agency, that can both be considered as applications of Bayesian inference to the problem of action selection: one that appears when trading off accuracy and uncertainty based on a general maximum entropy principle, and one that formulates action selection in terms of minimizing an error measure that quantifies deviations of beliefs and policies from given reference models. The first approach provides a normative rule for action selection in the face of model uncertainty or when information processing capabilities are limited. The second approach directly aims to formulate the action selection problem as an inference problem in the context of Bayesian brain theories, also known as Active Inference in the literature. We elucidate the main ideas and discuss critical technical and conceptual issues revolving around these two notions of free energy that both claim to apply at all levels of decision-making, from the high-level deliberation of reasoning down to the low-level information processing of perception.  

\textbf{Keywords:} free energy, intelligent agency, bayesian inference, maximum entropy, utility theory, active inference
\end{quote}
    
    \vspace{50pt}
  \end{@twocolumnfalse}
]

\thanks{This is a preprint of the article \href{https://doi.org/10.1371/journal.pcbi.100842}{\textit{The two kinds of free energy and the Bayesian revolution}, PLOS Computational Biology 16(12), 2020}.}

\section{Introduction}

There is a surprising line of thought connecting some of the greatest scientists of the last centuries, including Immanuel Kant, Hermann von Helmholtz, Ludwig E. Boltzmann, and Claude E. Shannon, whereby model-based processes of action, perception, and communication are explained with concepts borrowed from statistical physics. Inspired by Kant's Copernican revolution and motivated from his own studies of the physiology of the sensory system, Helmholtz was one of the first proponents of the \textit{analysis-by-synthesis} approach to perception \citep{Yuille2006}, whereby a perceiver is not simply conceptualized as some kind of \emph{tabula rasa} recording raw external stimuli, but rather relies on internal models of the world to match and anticipate sensory inputs. The internal model paradigm is now ubiquitous in the cognitive and neural sciences and has even led some researchers to propose a Bayesian brain hypothesis, whereby the brain would essentially be a prediction and inference engine based on internal models \citep{Kawato1999,Flanagan2003,Doya2007}.
Coincidentally, Helmholtz also invented the notion of the \textit{Helmholtz free energy} that plays an important role in thermodynamics and statistical mechanics, even though he never made a connection between the two concepts in his lifetime.

This connection was first made by Dayan, Hinton, Neal, and Zemel in their computational model of  perceptual processing as a statistical inference engine known as the \textit{Helmholtz machine} \citep{Hinton1995}. In this neural network architecture, there are feed-forward and feedback pathways, where the bottom-up pathway translates inputs from the bottom layer into hidden causes at the upper layer (the recognition model), and top-down activation translates simulated hidden causes into simulated inputs (the generative model). When considering log-likelihood in this setup as energy in analogy to statistical mechanics, learning becomes a relaxation process that can be described by the minimization of variational free energy. While it should be emphasized that variational free energy is not the same as Helmholtz free energy, the two free energy concepts can be formally related.
Importantly, variational free energy minimization is not only a hallmark of the Helmholtz machine, but of a more general family of inference algorithms, such as the popular expectation-maximization (EM) algorithm \citep{Neal1998,beal2003}. In fact, over the last two decades, variational Bayesian methods have become one of the foremost approximation schemes for tractable inference in the machine learning literature. 
Moreover, a plethora of machine learning approaches use loss functions that have the shape of a free energy when optimizing performance under entropy regularization in order to boost generalization of learning models \citep{Williams1991,Mnih2016}. 

In the meanwhile, free energy concepts have also made their way into the behavioral sciences. In the economic literature, for example, trade-offs between utility and entropic uncertainty measures that take the form of free energies have been proposed to describe decision-makers with stochastic choice behavior due to limited resources \citep{McKelvey1995,Sims2003,Mattsson2002,McFadden2005,Wolpert2006} or robust decision-makers with limited precision in their models \citep{Maccheroni2006,Hansen2008}. The free energy trade-off between entropy and reward can also be found in information-theoretic models of biological perception-action systems \citep{Still2009,Tishby2011,Ortega2013}, some of which have been subjected to 
experimental testing \citep{Ortega2016a,Sims2016,Schach2018,LindigLeon2019,Gershman2018,Griffiths2020}. Finally, in the neuroscience literature the notion of free energy has risen to recent fame as the central puzzle piece in the Free Energy Principle \citep{Friston2010} that has been used to explain a cornucopia of experimental findings including neural prediction error signals \citep{Sales2019}, synaptic plasticity rules\citep{Bogacz2017}, neural effects of biased competition and attention \citep{Friston2012,Parr2017}, visual exploration in humans \citep{Mirza2018}, and more---see the references in \citep{Friston2019}. Over time, the Free Energy Principle has grown out of an application of the free energy concept used in the Helmholtz machine, to interpret cortical responses in the context of predictive coding \citep{Friston2005}, and has gradually developed into a general principle for intelligent agency, also known as \emph{Active Inference} \citep{Friston2013,Friston2015,Friston2019}. Consequences and implications of the Free Energy Principle are discussed in neighbouring fields like psychiatry \citep{Schwartenbeck2016,Linson2020} and the philosophy of mind \citep{Clark2013,Colombo2018}. 

Given that the notion of free energy has become such a pervasive concept that cuts through multiple disciplines, the main rationale for this discussion paper is to trace back and to clarify different notions of free energy, to see how they are related and what role they play in explaining behavior and neural activity. As the notion of free energy mainly appears in the context of statistical models of cognition, the language of probabilistic models constitutes a common framework in the following discussion. Section 2 therefore starts with preliminary remarks on probabilistic modelling. Section 3 introduces two notions of free energy that are subsequently expounded in Section 4 and Section~5, where they are applied to models of intelligent agency. Section~6 concludes the paper.

\begin{figure}
\centering
\includegraphics[width=.5\textwidth]{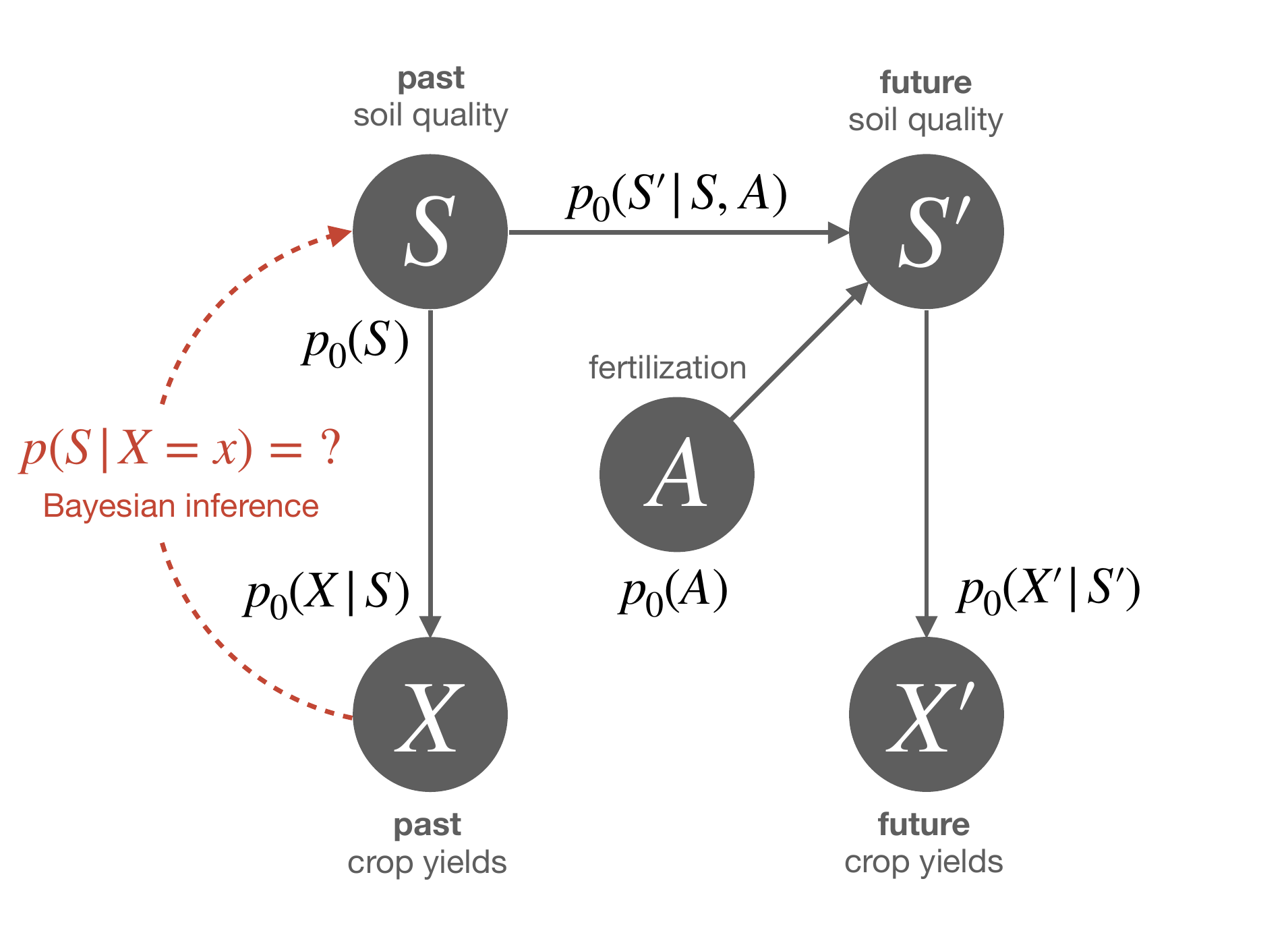}

\vspace{-10pt}
\caption{Graphical representation of an exemplary probabilistic model. 
The arrows (edges) indicate causal relationships between the random variables (nodes). The full joint distribution $p_0$ over all random variables is sometimes also referred to as a \emph{generative model}, because it contains the complete knowledge about the random variables and their dependencies and therefore allows to generate simulated data. Such a model could for example be used by a farmer to infer the soil quality $S$ based on the crop yields $X$ through Bayesian inference, which allows to determine a priori unknown distributions such as $p(S|X)$ from the generative model $p_0$ via marginalization and conditionalization.}
\label{fig:graphmodel}
\end{figure}   

\section{Probabilistic models and perception-action systems} \label{sec:probmodel}
Systems that show stochastic behavior, for example due to randomly behaving components or because the observer ignores certain degrees of freedom, are modelled using probability distributions. This way, any behavioral, environmental, and hidden variables can be related by their statistics, and dynamical changes can be modelled by changes in their distributions.

Consider, for example, the simple probabilistic model illustrated in Fig~\ref{fig:graphmodel}, consisting of the (for simplicity, discrete) variables \textit{past} and \textit{future soil quality} $\mathbf{S}\coloneqq (S,S')$, \textit{past} and \textit{future crop yields} $\mathbf{X}\coloneqq (X,X')$, and \textit{fertilization} $A$. The graphical model shown in the figure corresponds to the joint probability $p_0(\mathbf{X},\mathbf{S},A)$ given by the factorization
\begin{equation} \label{eq:exampleJoint}
p_0(X'|S') \, p_0(X|S)\, p_0(S'|S,A) \, p_0(S)  \, p_0(A)\, ,
\end{equation}
where $p_0(S)$ is the base probability of the past soil quality $S$, $p_0(X|S)$ is the probability of crop yields $X$ depending on the past soil quality $S$, and so forth.
Given the joint distribution we can now ask questions about each of the variables. For example, we could ask about the probability distribution $p(S|X\eqq x)$ of soil quality $S$ if we are told that the crop yields $X$ are equal to a value $x$. We can obtain the answer from the probabilistic model $p_0$ by doing \emph{Bayesian inference}, yielding the \textit{Bayes' posterior}
\begin{equation} \label{eq:bayes-posterior} 
p(S|X) = \frac{p(S,X)}{\sum_s p(s,X)} = \frac{p_0(X|S)p_0(S)}{\sum_s p_0(X|s)p_0(s)} \, ,
\end{equation}
where the dependencies on $X'$, $S'$, and $A$ have been summed out to calculate the marginal $p(S,X)$. In general, Bayesian inference in a probabilistic model means to determine the probability of some queried unobserved variables given the knowledge of some observed variables. This can be viewed as transforming the \textit{prior} probabilistic model $p_0$ to a \textit{posterior} model $p$, under which the observed values have probability one and unobserved variables have probabilities given by the corresponding Bayes' posteriors. 

In principle, Bayesian inference requires only two different kinds of operations, namely \textit{marginalization}, i.e.~summing out unobserved variables that have not been queried, such as $X',S'$ and $A$ above, and \textit{conditionalization}, i.e.~renormalizing the joint distribution over observed and queried variables---that may itself be the result from a previous marginalization such as $p(S,X)$ above---to obtain the required conditional distribution over the queried variables.
In practice, however, inference is a hard computational problem and many more efficient inference methods are available that may provide approximate solutions to the exact Bayes' posteriors, including belief propagation \citep{Pearl1988}, expectation propagation \citep{Minka2001}, variational Bayesian inference \citep{Hinton1993}, and Monte Carlo algorithms \citep{MacKay2002}. Also note that inference is trivial if the sought-after conditional distribution of the queried variable is already given by one of the conditional distributions that jointly specify the probabilistic model, e.g., $p(X|S) = p_0(X|S)$.

Probabilistic models can be used not only as \emph{external (observer) models}, but also as \emph{internal models} that are employed by the agent itself, or by a designer of the agent, in order to determine a desired course of action. In this latter case, actions could either be thought of as deterministic parameters of the probabilistic model that influence the future (\textit{influence diagrams}) or as random variables that are part of the probabilistic model themselves (\textit{prior models}) \citep{Boutilier1999}.  Either way, internal models allow making predictions over future consequences in order to find actions or distributions over actions that lead to desirable outcomes, for example actions that produce high rewards in the future. In \textit{mechanistic} or \textit{process} model interpretations, some of the specification procedures to find such actions are themselves meant to represent what the agent is actually doing while reasoning, whereas \textit{as if} interpretations simply use these methods as tools to arrive at distributions that describe the agent's behavior. Free energy is one of the concepts that appears in both types of methods.

\section{The two notions of free energy} \label{sec:freeenergy}

Vaguely speaking, free energy can refer to any quantity that is of the form
\begin{equation}\label{free-energy}
\text{free energy}\ = \  \text{energy} \, \pm \, \text{const.} \times  \text{entropy}  ,
\end{equation}
where \textit{energy} is an expected value of some quantitity of interest, \textit{entropy} refers to a quantity measuring disorder, uncertainty, or complexity, that must be specified in the given context, and \textit{const.} is a constant term that translates between units of entropy and energy, and is related to the \textit{temperature} in physically motivated free energy expressions. From relation \eqref{free-energy}, it is not surprising that free energy sometimes appears enshrouded by mystery, as it relies on an understanding of entropy, and ``nobody really knows what entropy is anyway'', as John Von Neumann famously quipped \citep{Feynman1996}.

Historically, the concept of free energy goes back to the roots of thermodynamics, where it was introduced to measure the maximum amount of work that can be extracted from a thermodynamic system at a constant temperature and volume.  If, for example, all the molecules in a box move to the left, we can use this kinetic energy to drive a turbine. If, however, the same kinetic energy is distributed as random molecular motion, it cannot be fully transformed into work. Therefore,  only part of the total energy $E$ is usable, because the exact positions and momenta of the molecules, the so-called \textit{microstates}, are unknown. In this case, the maximum usable part of the energy $E$ is the \emph{Helmholtz free energy}, defined as 
\begin{equation} \label{Helmholtz-freeenergy}
F = E - T S \, ,
\end{equation}
where $S$ is the thermodynamic entropy. In general, the transformation between two macrostates with free energies $F_1$ and $F_2$ allows the extraction of work $W\leq F_2-F_1$.

While the two notions of free energy that we discuss in the following are vaguely inspired by the physical original, their motivations are rather distinct and the main reason they share the nomenclature is due to their general form \eqref{free-energy} resembling the Helmholtz free energy \eqref{Helmholtz-freeenergy}.

\subsection{Free energy from constraints} \label{sec:maxent}

The first notion of free energy is closely tied to the \textit{principle of maximum entropy} \citep{Jaynes1957}, which virtually appears in all branches of science. From this vantage point, the physical free energy is merely a special instance of a more general inference problem where we hold probabilistic beliefs about unknown quantities (e.g., the exact energy values of the molecules in a gas) and we can only make coarse measurements or observations (e.g., the temperature of the gas) that we can use to update our beliefs about these hidden variables. The principle of maximum entropy suggests that, among the beliefs that are compatible with the observation, we should choose the most ``unbiased'' belief, in the sense that it corresponds to a maximum number of possible assignments of the hidden variables.

\subsubsection{Wallis' motivation of the maximum entropy principle}
Consider the random experiment of distributing $N$ elements randomly in $n$ equally probable buckets with $N\gg n$, where the resulting number of elements $N_i$ in bucket $i\in \{1,\dots,N\}$ determines the probability $p(z_i)\coloneqq \frac{N_i}{N}$. In principle, this way we could generate any distribution $p$ over a finite set $\Omega=\{z_1,\dots,z_n\}$ that we like, however, a uniform distribution that reflects the equiprobable assignment clearly is much more likely than a Dirac distribution where all the probability mass is concentrated in one bucket. Here, the reason is that there are many possible assignments of elements among the buckets that generate the uniform distribution, whereas there is only one for a Dirac distribution. In fact, the number of possibilities of how to distribute $N$ elements among $n$ buckets with $N_i$ elements in the $i$th bucket is
\begin{equation} \label{eq:chooseNN}
\omega \coloneqq \frac{N!}{N_1!\cdots N_n!} \, ,
\end{equation}
because $N!$ is the number of possible permutations of all $N$ elements, which overcounts by the number of permutations of elements inside the same bucket and thus has to be divided by the number of permutations $N_i!$ for all $i=1,\dots,n$. In the absence of any further measurement constraints, the number of possibilities \eqref{eq:chooseNN} is maximized by $N_i=N/n$ for all $i$, and thus the \textit{typical} distribution $p^\ast$ over $\Omega$ in this case is the uniform distribution, i.e., $p^\ast(z_i)=\frac{1}{n}$ for all $i$.  

Consider now the problem of having to determine a typical distribution $p^\ast$ over $\Omega$ such that the expected value $\mathbb E_{p^\ast}[\mathcal E] \eqqcolon \langle \mathcal E\rangle_{p^\ast}$ of some quantity $\mathcal E$ equals a measured value $\varepsilon$. A simple example would be the experiment of throwing $N$ dice and taking $\mathcal E$ to be the number of dots, i.e., $\mathcal E(z_1)=1,\dots,\mathcal E(z_6)=6$, and trying to find the typical distribution $p^\ast$ over outcomes $z_1,\dots,z_6$ under the constraint that the average number of dots is, say $\varepsilon = 2$. The solution to this problem is analogous to the case of no constraints, but this time we only consider realizations that are compatible with the measurement constraint, that is we let $(N_1,\dots,N_n)$ belong to the set of permissible occupation vectors  
\[
\Gamma_\varepsilon \coloneqq \big\{(N_1,\dots,N_n) \, \big | \, \langle \mathcal E \rangle_p = \varepsilon, \ p(z_i)=\tfrac{N_i}{N} \ \forall i\big\} .
\]
A typical distribution $p^\ast$ for a constraint $\varepsilon$ can then be determined by a candidate in $\Gamma_\varepsilon$ with the maximum number $\omega$ of possibilities \eqref{eq:chooseNN}. By assumption, $N$ is much larger than $n$, so that we can get rid of the faculties by making use of Stirling's approximation $\ln N! = N\ln N - N + \mathcal O(\ln N)$. In particular, when letting $N,N_i\to\infty$ such that $p(z_i)=\frac{N_i}{N}$ remains finite, we obtain
\[
\frac{1}{N} \log \omega  = \underbrace{-\sum_{i=1}^n \frac{N_i}{N}\log \frac{N_i}{N}}_{=H(p)} + \, \mathcal O\left(\frac{\log N}{N}\right) \ \stackrel{N\to\infty}{\longrightarrow}  \ H(p) \, .
\]
where $H(p)\coloneqq -\sum_{z\in\Omega} p(z) \log p(z)$ denotes the (\textit{Gibbs} or \textit{Shannon}) \textit{entropy} of $p$. Thus, instead of assessing typicality by maximizing \eqref{eq:chooseNN} in $\Gamma_\varepsilon$ for large but fixed $N$, we can get rid of the $N$-dependency by simply maximizing $H$, 
\begin{equation} \label{eq:maxEnt}
p^\ast = \argmax_{p, \langle \mathcal E \rangle_p = \varepsilon} \ H(p) \, .
\end{equation}
This constrained optimization problem is known as the principle of maximum entropy. The motivation given here is essentially the Wallis derivation presented by Jaynes \citep{Jaynes2003}.

\subsubsection{Free energy from constraints and the Boltzmann distribution}
The constrained optimization problem \eqref{eq:maxEnt} can be translated into an unconstrained problem by introducing a Lagrange multiplier $\beta$, known as the \textit{inverse temperature} due to the analogy to thermodynamics and the Helmholtz Free Energy \eqref{Helmholtz-freeenergy}, which has to be chosen post hoc such that the constraint is satisfied. This results in the minimization of the Lagrangian 
\begin{equation}\label{eq:constraintFE1}
F(p) \coloneqq \langle \mathcal E \rangle_p \, {-}\, \tfrac{1}{\beta} H(p) ,
\end{equation}
which takes the form of a free energy \eqref{free-energy}. As we shall see later, $F$ takes its minimum at the \textit{Boltzmann distribution} known from statistical mechanics, given by
\begin{equation}\label{def:thermodyn:boltzmann}
p^\ast (z) \coloneqq \frac{1}{\mathcal Z} \, e^{- \beta \mathcal E(z)},
\end{equation}
where $\mathcal{Z} = \sum_{z\in\Omega} e^{-\beta \mathcal E(z)}$ denotes the normalization constant.

Note that, the argument in the previous section implicitly assumes a \textit{uniform} reference distribution, because the buckets are assumed to be equiprobable. When replacing this assumption by the assumption of a general distribution $p_0$ over $\Omega$, we obtain the \textit{principle of minimum relative entropy} \citep{Jaynes1983}, where the so-called \textit{Kullback-Leibler} (KL) \textit{divergence} $\DKL(p\|p_0)=\langle\log (p/p_0) \rangle_p$ is minimized with respect to $p$ subject to a constraint $\langle \mathcal E\rangle_p = \varepsilon$. Analogous to the maximum entropy principle, this translates to the unconstrained minimization of the Lagrangian
\begin{equation} \label{eq:constraintFE2}
F(p,p_0) \coloneqq \langle \mathcal E \rangle_p \, {+}\, \tfrac{1}{\beta} D_\mathrm{KL}(p\|p_0),
\end{equation}
with solution given by $p^\ast(z) = \frac{1}{\mathcal Z} \, p_0(z) \, e^{-\beta \mathcal E(z)}$.

\subsubsection{The trade-off between energy and uncertainty}

\begin{figure}
\centering
\includegraphics[width=.5\textwidth]{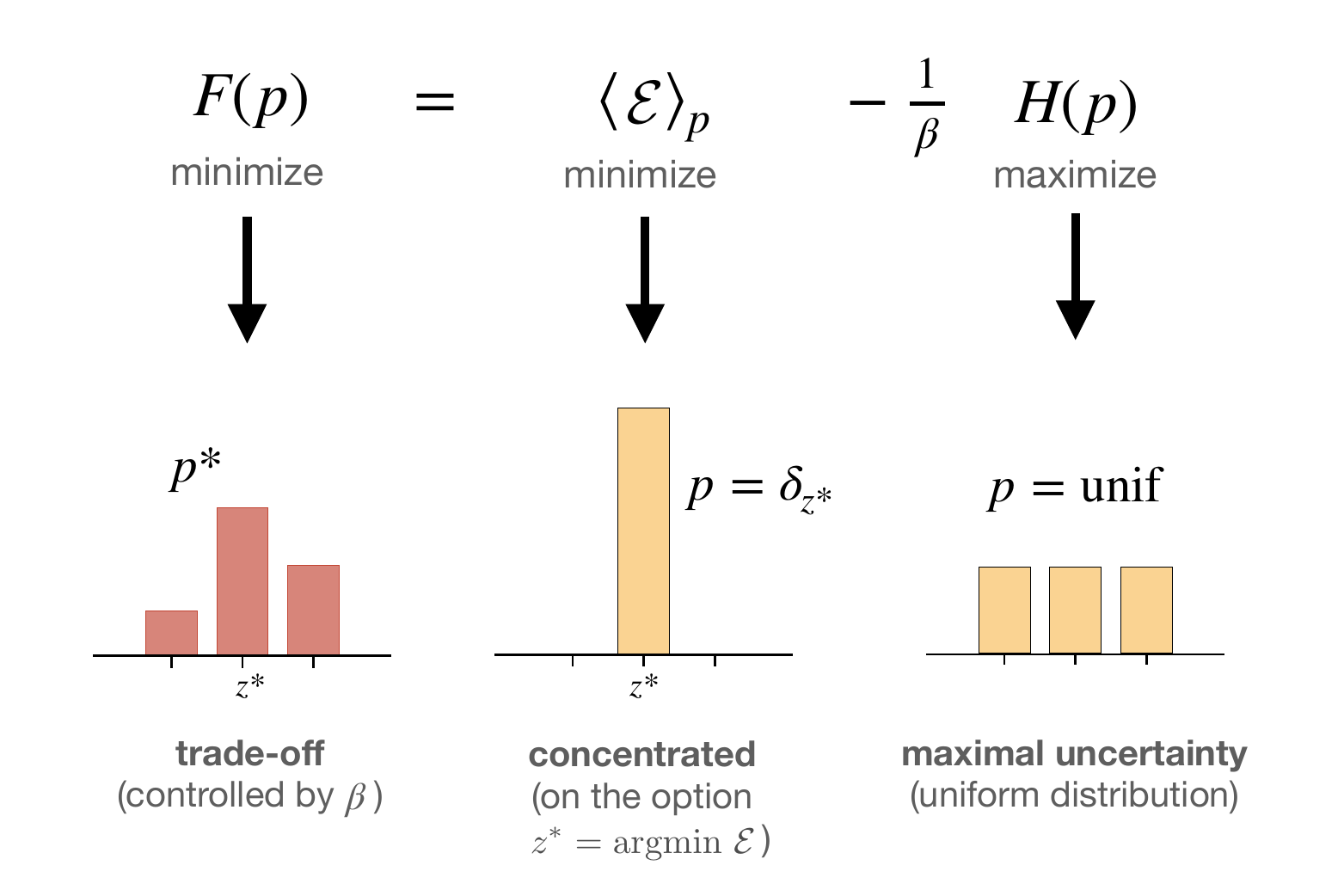}

\vspace{-10pt}
\caption{Minimizing the free energy from constraints \eqref{eq:constraintFE1} requires to trade off the competing terms of energy $\langle \mathcal E \rangle_p$ and entropy $H(p)$, here shown exemplarily for the case of three elements. Assuming there exists a unique minimal element $z^\ast=\argmin_z \mathcal E(z)$, then minimizing only $\langle \mathcal E \rangle_p$ over all probability distributions $p$ results in the (\textit{Dirac delta}) distribution $\delta_{z^\ast}$ that assigns zero probability to all $z_i\not = z^\ast$ and probability one to $z_i \eeqq z^\ast$, and therefore has zero entropy. In contrast, minimizing only the term $-H(p)/\beta$ is equivalent to maximizing $H(p)$ and therefore would result in the uniform distribution that gives equal probability to all elements. The resulting Boltzmann distribution $p^\ast$ interpolates between these two extreme solutions of minimal energy ($\beta\to\infty$) and maximum entropy ($\beta\to 0$).}
\label{fig:trade-off}
\end{figure}   

An important feature of the minimization of the free energies \eqref{eq:constraintFE1} and \eqref{eq:constraintFE2} consists in the balancing of the two competing terms of energy and entropy (cf. Fig \ref{fig:trade-off}). This \textit{trade-off} between maximal uncertainty (uniform distribution, or $p_0$) on the one hand and minimal energy (e.g.~a delta distribution) on the other hand is the core of the maximum entropy principle. The inverse temperature $\beta$ plays the role of a trade-off parameter that controls how these two counteracting forces are weighted.

The maximum entropy principle goes back to the \textit{principle of insufficient reason} \citep{Bernoulli1713,Laplace1812,Poincare1912}, which states that two events should be assigned the same probability if there is no reason to think otherwise. It has been hailed as a principled method to determine prior distributions and to incorporate novel information into existing probabilistic knowledge. In fact, Bayesian inference can be cast in terms of relative entropy minimization with constraints given by the available information \citep{Williams1980}. Applications of this idea can also be found in the machine learning literature, where subtracting (or adding) an entropy term from an expected value of a function that must be optimized is known as \textit{entropy regularization} and plays an important role in modern reinforcement learning algorithms \citep{Williams1991,Mnih2016} to encourage exploration \citep{Levine2017} as well as to penalize overly deterministic policies resulting in biased reward estimates \citep{Tishby2016}. 

From now on, we refer to a free energy expression that is motivated from a trade-off between an energy and an entropy term, such as \eqref{eq:constraintFE1} and \eqref{eq:constraintFE2}, as \textit{free energy from constraints}, in order to discriminate it from the notion of free energy introduced in the following section, which---despite of its resemblance---has a different motivation.

\medskip

\subsection{Variational free energy} \label{sec:varFE}

There is another, distinct appearance of the term ``free energy'' outside of physics, which is a priori \textit{not} motivated from a trade-off between an energy and entropy term, but from 
possible efficiency gains when representing Bayes' rule in terms of an optimization problem. This technique is mainly used in \textit{variational Bayesian inference} \citep[Ch.~11]{Koller2009}, originally introduced by Hinton and van Camp \citep{Hinton1993}.  As before, for simplicity all random variables are discrete, but most expressions can directly be translated to the continuous case by replacing probability distributions by probability densities and sums by the corresponding integrals.

\begin{figure*}
\includegraphics[width=1\textwidth]{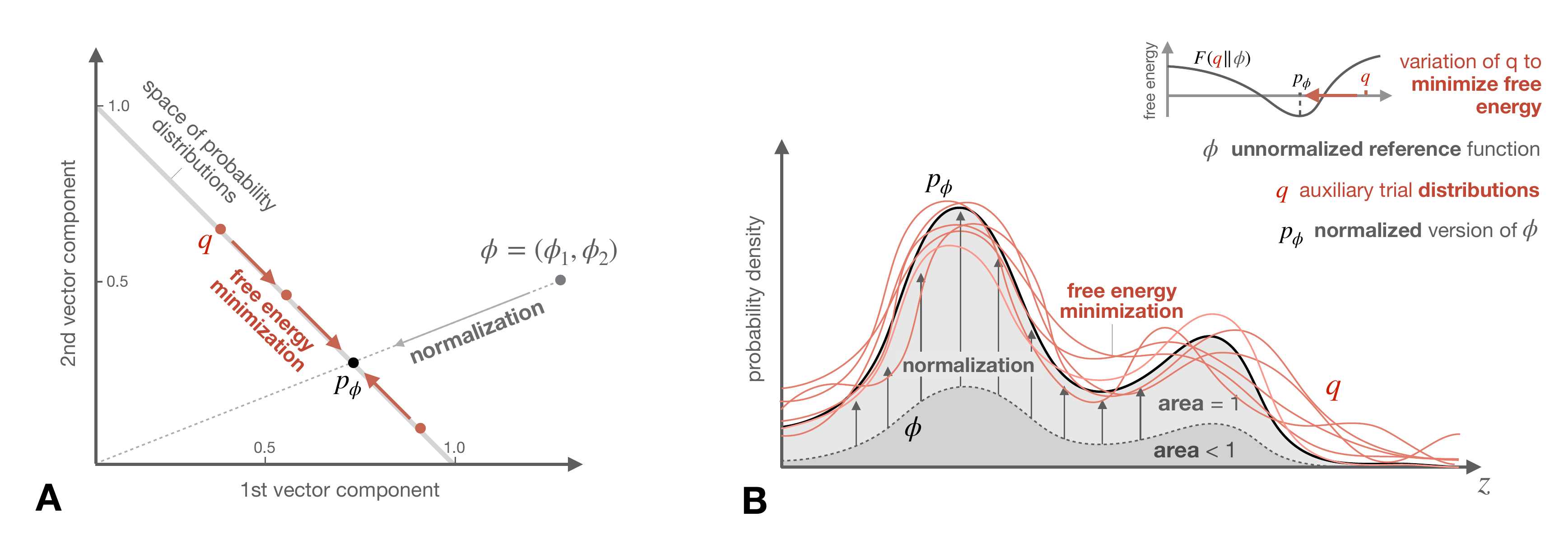}

\vspace{-10pt}
\caption{The normalization of a functon $\phi$ to obtain a probability distribution $p_\phi$ is equivalent to fitting trial distributions $q$ to the shape of $\phi$ by minimizing free energy. In two dimensions, the normalization of a point $\phi=(\phi_1,\phi_2)$ corresponds to a (non-orthogonal) projection onto the plane of probability vectors (\textbf{A}). For continuous domains, where probability distributions are represented by densities, normalization corresponds to a rescaling of $\phi$ such that the area below the graph equals $1$ (\textbf{B}). Instead, when minimizing variational free energy (red colour), the trial distributions $q$ are varied until they fit to the shape of the unnormalized function $\phi$ (perfectly at $q=p_\phi$).}
\label{fig:normalization}
\end{figure*} 

\subsubsection{Variational Bayesian inference} \label{sec:variationalinference}

As we have seen in Section \ref{sec:probmodel}, Bayesian inference consists in the calculation of a conditional probability distribution over unknown variables given the values of known variables. In the most simple case of two variables, say $X$ and $Z$, and a probabilistic model of the form $p_0(X,Z) = p_0(X|Z)p_0(Z)$, Bayesian inference applies if $X$ is observed and $Z$ is queried. Analogous to \eqref{eq:bayes-posterior}, the \textit{exact} Bayes' posterior $p(Z|X\eqq x)$ is defined by the renormalization of $p_0(x,Z)$ in order to obtain a distribution over $Z$ that respects the new information $X\eqq x$,
\begin{equation} \label{eq:exactBayesVar}
p(Z|X\eqq x) \, = \, \frac{p_0(x,Z)}{\mathcal Z(x)} \,  \, = \, \frac{p_0(x|Z) \, p_0(Z)}{\mathcal Z(x)} \,  \, ,
\end{equation}
with the normalization constant $\mathcal Z(x) = \sum_z p_0(x,z)$.

In \textit{variational} Bayesian inference, however, this Bayes' posterior is not calculated directly by renormalizing $p_0(x,Z)$ with respect to $Z$, but indirectly by approximating it by a distribution $q(Z)$ that is adjusted through the minimization of an error measure that quantifies the deviation from the exact Bayes' posterior. Importantly, the value of this error measure can be determined without having to know the exact Bayes' posterior. To see this, note that the KL divergence between $q(Z)$ and $p(Z|X\eqq x)$ can be written as
\begin{equation} \label{eq:variationalFEderivation}
\underbrace{\left\langle \log \frac{q(Z)}{p(Z|X\eqq x)} \right\rangle_{q(Z)}}_{= \, D_\mathrm{KL}(q(Z)\|p(Z|X\eqq x))} \ = \  \underbrace{\log \mathcal Z(x)}_{\text{indep.~of } q} + \underbrace{\left\langle \log \frac{q(Z)}{p_0(x,Z)} \right\rangle_{q(Z)}}_{\eqqcolon \, F(q(Z)\|p_0(x,Z))} \, ,
\end{equation}
i.e., it can be decomposed into the sum of a constant term and a term that does not depend on the normalization $\mathcal Z(x)$. In particular, a good approximation $q(Z)$ of the exact Bayes' posterior \eqref{eq:exactBayesVar} will effectively minimize this KL divergence, which---due to \eqref{eq:variationalFEderivation}---can be done by minimizing $F(q(Z)\|p_0(x,Z))$. In particular, the optimium of this minimization is exactly achieved at the Bayes' posterior \eqref{eq:exactBayesVar},
\begin{equation}\label{eq:varbayes}
 \argmin_{q(Z)} \left\langle \log \frac{q(Z)}{p_0(x,Z)} \right\rangle_{\hspace{-2pt} q(Z)} = \ p(Z|X\eqq x) \, \, ,
\end{equation}
which is known as the variational characterization of Bayes' rule. This result is a special case of \eqref{eq:argminF} in the following section.

\begin{figure*}
\includegraphics[width=1\textwidth]{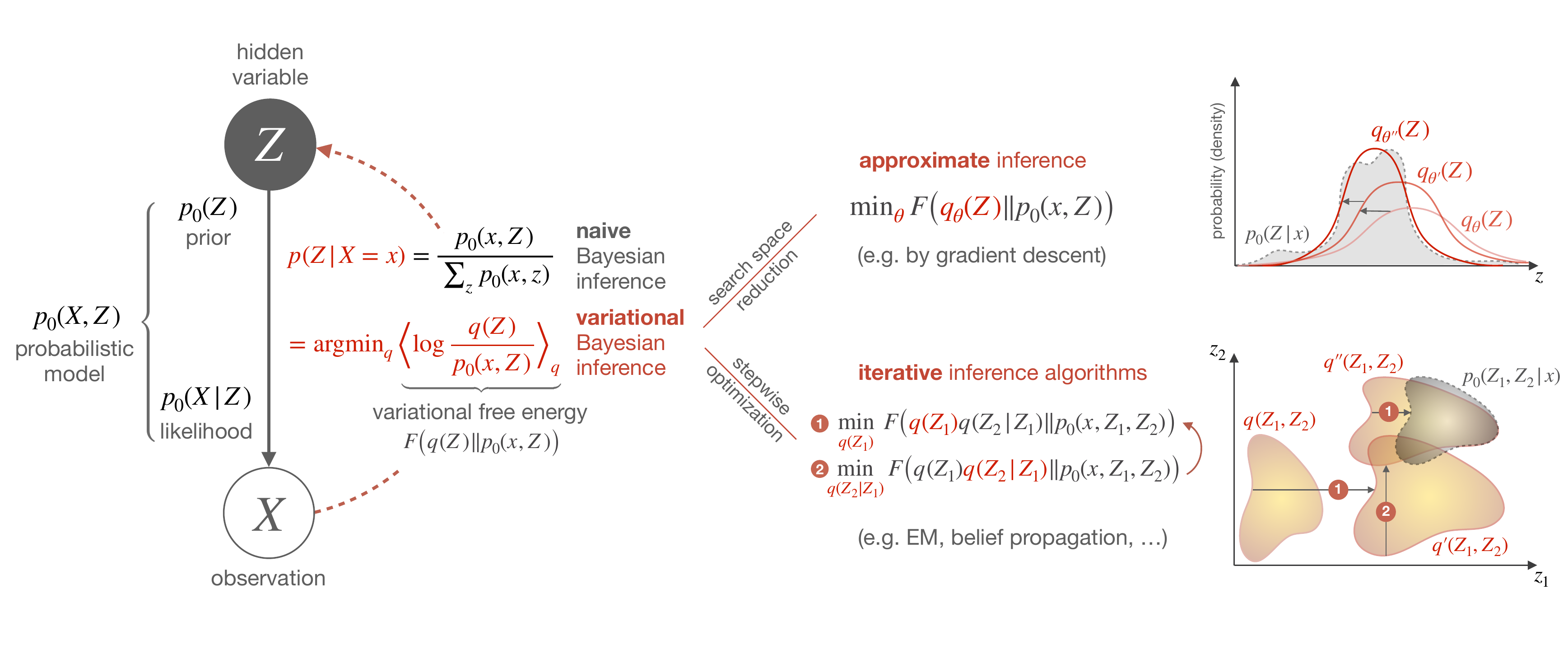}

\vspace{-10pt}
\caption{In variational Bayesian inference, the operation of renormalizing the probabilistic model $p_0$ evaluated at an observation $X\eqq x$ (Bayes' rule), is replaced by an optimization problem. In practice, this variational representation is often exploited to simplify a given inference problem, either by reducing the seach space of distributions, for example through a restrictive parametrization resulting in approximate inference, or by splitting up the optimization into multiple partial optimization steps that are potentially easier to solve than the original problem but might still converge to the exact solution. These two simplifications can also be combined, for example in the case of mean-field assumptions where the space of distributions is reduced \textit{and} an efficient iterative inference algorithm is obtained at the same time.}
\label{fig:bayesianinference}
\end{figure*}   

\subsubsection{Variational free energy, an extension of relative entropy}
\label{sec:normalization}

Any non-negative function $\phi$ on a finite space $\Omega$, can be normalized to obtain a probability distribution $p_\phi=\phi/\sum_z \phi(z)$ on $\Omega$ that differs from $\phi$ only by a scaling constant. In cases when it is not beneficial to carry out the sum $\sum_z \phi(z)$ explicitly, such a normalization might be replaced by the minimization of the \textit{variational free energy}
\begin{equation}\label{def:var-free-energy}
F(q\|\phi) \coloneqq \left \langle \log \frac{q(Z)}{\phi(Z)} \right\rangle_{\hspace{-2pt}q(Z)} ,
\end{equation}
with respect to the so-called \textit{trial distributions} $q$, because we have
\begin{align}  \label{eq:argminF}
\argmin_q F(q\|\phi)  & \ = \  \frac{\phi(Z)}{\sum_z \phi(z)}  \ = \ p_\phi(Z) \, .
\end{align}
Thus, instead of normalizing $\phi$ directly, one fits auxiliary distributions $q$ to approximate the shape of $\phi$ in the space of probability distributions (cf.~Fig \ref{fig:normalization}). If this optimization process has no constraints, then the trial distributions are adjusted until $p_\phi$ is achieved. In the case of constraints, for instance if the trial distributions are parametrized by a non-exhaustive parametrization (e.g., Gaussians), then the optimized trial distributions approximate $p_\phi$ as close as possible within this parametrization. The minimal value of $F(q\|\phi)$ is
\begin{equation}  \label{eq:minF}
F(p_\phi\|\phi) \, = \, \min_q F(q\|\phi)  \, = \,  -\log \sum_z \phi(z) \, .
\end{equation}
In particular, this implies that $ -F(q\|\phi)  \leq  \log \sum_z \phi(z)$ for all $q$, so that varying $-F(q\|\phi)$ with arbitrary trial distributions $q$ always provides a lower bound to the unknown normalization constant $\sum_z \phi(z)$. In Bayesian inference this is the normalization constant in Bayes' rule and called the \textit{model evidence}, which is why the negative variational free energy is also called \textit{evidence lower bound} (ELBO). 

The proof of \eqref{eq:argminF} and \eqref{eq:minF} directly follows from Jensen's inequality and only relies on the concavity of the logarithm. As we have seen in the previous section, in variational Bayesian inference, the \textit{reference} $\phi$ usually takes the form of a joint distribution evaluated at the observed variables, e.g., $\phi(Z) = p_0(x,Z)$ in which case \eqref{eq:argminF} recovers \eqref{eq:varbayes}. The variational free energy \eqref{def:var-free-energy} is a free energy in the sense of \eqref{free-energy} since by the additivity of the logarithm under multiplication ($\log ab=\log a +\log b$),
\begin{equation} \label{variationalIsFreeEnergy}
F(q\|\phi) = \langle -\log\phi\rangle_{q} - H(q)
\end{equation}
with energy term $\langle-\log \phi\rangle_q$ and entropy term $H(q)$. Note that, for the choice $\phi = e^{-\beta \mathcal E}$, Equation \eqref{eq:argminF} becomes the Boltzmann distribution \eqref{def:thermodyn:boltzmann} and the variational free energy \eqref{variationalIsFreeEnergy} formally corresponds to the free energy from constraints \eqref{eq:constraintFE1}.

Variational free energy can be regarded as an extension of relative entropy with the reference distribution being replaced by a non-normalized reference function, since in the case when $\phi$ is already normalized, that is if $\sum_z \phi(z) = 1$, then the free energy \eqref{def:var-free-energy} coincides with the KL divergence $D_\mathrm{KL}(q\|\phi)$. In particular, while relative entropy is a measure for the dissimilarity of two probability distributions, where the minimum is achieved if both distributions are equal, variational free energy is a measure for the dissimilarity between a probability distribution $q$ and a (generally non-normalized) function $\phi$, where the minimum with respect to $q$ is achieved at $p_\phi$. Accordingly, we can think of the variational free energy as a specific error measure between probability distributions and reference functions. In principle, one could design many other error measures that have the same minimum. This means that, a statement in a probabilistic setting that a distribution $q^\ast$ minimizes a variational free energy $F(q\|\phi)$ with respect to a given reference $\phi$, is analogous to a statement in a non-probabilistic setting that some number $x=x^\ast$ minimizes the value of an error measure $\epsilon(x,y)$ (e.g., the squared error $\epsilon(x,y) = (x-y)^2$) with respect to a given reference value $y$.

\subsubsection{Approximate and iterative inference} \label{sec:approxIterativeInf}
Representing Bayes' rule as an optimization problem over auxiliary distributions $q$ has two main applications that both can simplify the inference process (cf. Fig \ref{fig:bayesianinference}). First, it allows to \textit{approximate} exact Bayes' posteriors by restricting the optimization space, for example using a non-exhaustive parametrization, e.g.~an exponential family. Second, it enables \textit{iterative inference algorithms} consisting of multiple simpler optimization steps, for example by optimizing with respect to each term in a factorized representation of $q$ separately. A popular choice is the \textit{mean-field approximation}, which combines both of these simplifications, as it assumes independence between hidden states, effectively reducing the search space from joint distributions to factorized ones, and moreover it allows to optimize with respect to each factor alternatingly. Note, however, that mean-field approximations have limited use in sequential environments, where independence of subsequent states cannot be assumed and therefore less restrictive assumptions must be used instead \citep{Opper2001}.

Many efficient iterative algorithms for exact and approximate inference can be viewed as examples of variational free energy minimization, for example the EM algorithm \citep{Dempster1977,Neal1998}, belief propagation \citep{Pearl1988,Yedidia2001}, and other message passing algorithms\citep{Minka2001,Wainwright2005,Winn2005,Minka2005,Yedidia2005}. While the (Bayesian) EM algorithm \citep{beal2003} and Pearl's belief propagation \citep{Yedidia2001} both can be seen as minimizing the same variational free energy, just with different assumptions on the approximate posteriors, in \citep{Minka2005}, it is shown that also many other message passing algorithms such as \citep{Minka2001,Wainwright2005,Winn2005} can be cast as minimizing some type of free energy, the only difference being the choice of the divergence measure as the entropy term. Simple versions of these algorithms have often existed before their free energy formulations were available, but the variational representations usually allowed for extensions and refinements---see \citep{Csiszar1984,Hathaway1986,Neal1998,beal2003} in case of EM and \citep{Yedidia2001,Heskes2002,Yuille2002,Yedidia2005} in case of message passing.

\medskip 

We are now turning to the question of how the two notions of free energy introduced in this section are related to recent theories of intelligent agency.

\section{Free energy from constraints in information processing} \label{sec:utility-based-normative}

\subsection{The basic idea}

The concept of free energy from constraints as a trade-off between energy and uncertainty can be used in models of perception-action systems, where entropy quantifies information processing complexity required for decision-making (e.g., planning a path for fleeing a predator) and energy corresponds to performance (e.g., distinguishing better and worse flight directions). The notion of decision in this context is very broad and can be applied to any internal variable in the perception-action pipeline \citep{Kahneman2002}, that is not given directly by the environment. In particular, it also subsumes perception itself, where the decision variables are given by the hidden causes that are being inferred from observations.

In rational choice theory \citep{Neumann1944}, a decision-maker selects decisions $x^\ast$ from a set of options $\Omega$ such that a utility function $U$ defined on $\Omega$ is maximized, 
\begin{equation}\label{eq:maxutil}
x^\ast = \argmax_{x\in\Omega} \, U(x) \, .
\end{equation}
The utility values $U(x)$ could either be objective, for example a monetary gain, or subjective in which case they represent the decision-maker's preferences. In general, the utility does not have to be defined directly on $\Omega$, but could be derived from utility values that are attached to certain states, for example to the configurations of the playboard in a board game. In the case of perception, utility values are usually given by (log-)likelihood functions, in which case utility maximization without constraints corresponds to greedy inference such as maximum likelihood estimation. Note that, for simplicity, in this section we consider one-step decision problems. Sequential tasks can either be seen as multiple one-step problems where the utility of a given step might depend on the policy over future steps, or as path planning problems where an action represents a full action path or policy \citep{Whittle1990,Tishby2011,GrauMoya2016,Gottwald2018}.

\begin{figure}
\centering
\includegraphics[width=0.45\textwidth]{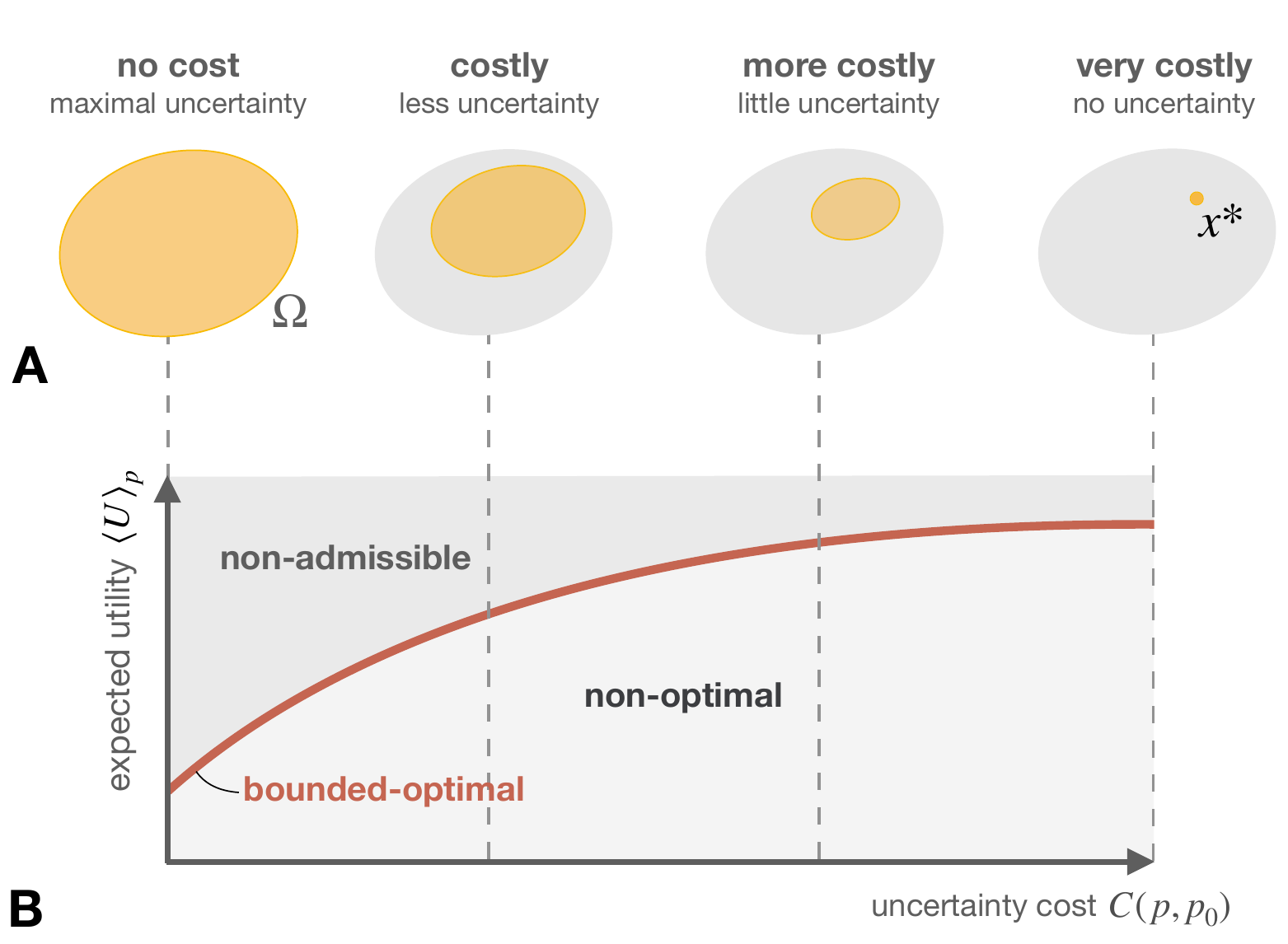}

\caption{\textbf{A}: Decision-making can be considered as a search process in the space of options $\Omega$, where options are progressively ruled out. Deliberation costs are defined to be monotone functions under such uncertainty reduction. \textbf{B}: Exemplary efficiency curve resulting from the trade-off between utility and costs, that separates non-optimal from non-admissible behavior. The points on the curve correspond to bounded-optimal agents that optimally trade off utility against uncertainty, analogous to the rate-distortion curve in information theory.}
\label{fig:boundedrat}
\end{figure}  

While ideal rational decision-makers are assumed to perfectly optimize a given utility function $U$, real behavior is often stochastic, meaning that multiple exposures to the same problem lead to different decisions. Such non-deterministic behavior could be a consequence of model uncertainty, as in Bayesian inference or various stochastic gambling schemes, or a consequence of \emph{satisficing} \citep{Simon1955}, where decision-makers do not choose the single best option, but simply one option that is good enough. Abstractly, this means that, the choice of a single decision is replaced by the choice of a distribution over decisions. More generally, also considering prior information that the decision-maker might have from previous experience, the process of deliberation during decision-making might be expressed as the transformation of a prior $p_0$ to a posterior distribution $p$. 

When assuming that deliberation has a cost $C(p,p_0)$, then arriving at narrow posterior distributions should intuitively be more costly than choosing distributions that contain more uncertainty (cf.~Fig \ref{fig:boundedrat}A). In other words, deliberation costs must be increasing with the amount of uncertainty that is reduced by the transformation from $p_0$ to $p$. Uncertainty reduction can be understood as making the probabilities of options less equal to each other, rigorously expressed by the mathematical concept of majorization \citep{Marshall2011}. This notion of uncertainty can also be generalized to include prior information, so that the degree of uncertainty reduction corresponds to more or less deviations from the prior \citep{Gottwald2019}.

Maximizing expected utility $\langle U\rangle_p$ with respect to $p$ under restrictions on processing costs $C(p,p_0)$ is a constrained optimization problem that can be interpreted as a particular model of \textit{bounded rationality} \citep{Simon1955}, explaining non-rational behavior of decision-makers that may be unable to select the single best option by their limited information processing capability. Similarly to the free energy trade-off between energy and entropy (cf. Fig \ref{fig:trade-off}), this results in a trade-off between utility $\langle U\rangle_p$ and processing costs $C(p,p_0)$,
\begin{equation} \label{eq:broundedrational-general}
F_\beta(p) \coloneqq \langle U\rangle_p  - \tfrac{1}{\beta} C(p,p_0).
\end{equation}
Here, the trade-off parameter $\beta$ is analogous to the \textit{inverse temperature} in statistical mechanics (cf.~Equation \eqref{eq:constraintFE1}) and parametrizes the optimal trade-offs $p_\beta^\ast \eeqq \argmax_{p} F_\beta(p)$ between utility and cost, that define an efficiency frontier separating the space of perception-action systems into bounded-optimal, non-optimal, and non-admissible systems (cf.~Fig \ref{fig:boundedrat}).

When assuming that the total transformation cost is the same independent of whether a decision problem is solved in one step or multiple sub-steps (\textit{additivity under coarse-graining}) the trade-off in \eqref{eq:broundedrational-general} takes the general form \eqref{free-energy} of a free energy in the sense of energy (utility) minus entropy (cost), because then the cost function is uniquely given by the relative entropy
\begin{equation}\label{eq:informationcost}
C(p,p_0) = \DKL(p\|p_0). 
\end{equation}
Note that the additivity of \eqref{eq:informationcost} also implies a coarse-graining property of the free energy \eqref{eq:broundedrational-general} in the case when the decision is split into multiple steps, such that the utility of preceding decisions is effectively given by the free energy of following decisions. Therefore, in this case, free energy can be seen as a \emph{certainty-equivalent} value of the subordinate decision problems, i.e.~the amount of utility the agent would have to receive to be indifferent between this guaranteed utility and the potential expected utility of the subsequent decision steps taking account the associated information processing costs. The special case \eqref{eq:informationcost} has been studied extensively in multiple contexts, including quantal response equilibria in the game-theoretic literature \citep{McKelvey1995,Wolpert2006}, rational inattention and costly contemplation \citep{Sims2003,Ergin2010}, bounded rationality with KL costs \citep{Mattsson2002,Ortega2013}, KL control \citep{Todorov2009,Kappen2012}, entropy regularization \citep{Williams1991,Mnih2016}, robustness \citep{Maccheroni2006,Hansen2008}, the emergence of heuristics \citep{Binz2020}, thermodynamic models of computation \citep{Wolpert2019}, and the analysis of information flow in perception-action systems \citep{Tishby2011,Still2009}. While  \eqref{eq:informationcost}  is often regarded as an abstract measure of uncertainty reduction or a generic proxy for information processing costs, it can also be viewed as a physical capacity constraint, where the information that is required to achieve a certain expected utility is considered to be sent over a channel to the actuator \citep{Miller1956,Garner1962,MacRae1970,TatikondaMitter2004,Gershman2018}. 
This view is also consistent with the maximum entropy principle, as \eqref{eq:broundedrational-general} and~\eqref{eq:informationcost} favor distributions $p$ that can be generated from $p_0$ most easily in terms of statistics, and therefore with minimum communication complexity between $p_0$ and $p$ \citep{Harsha2009}.

\begin{figure}
\centering
\includegraphics[width=.48\textwidth]{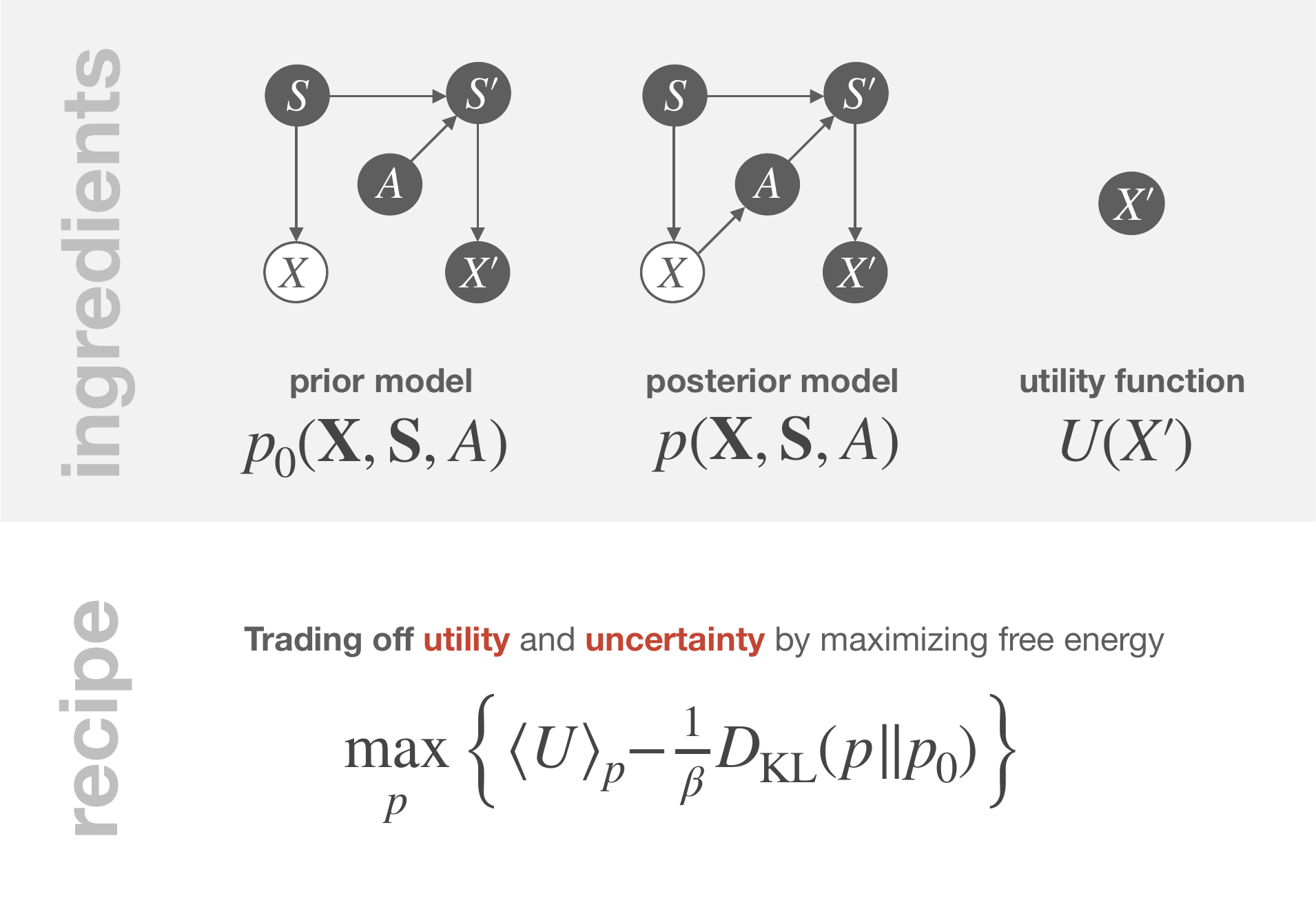}

\vspace{-10pt}
\caption{Overview of how to apply utility maximization with information processing costs to the example from Section \ref{sec:probmodel}.}
\label{fig:boundedRatExample}
\end{figure}  

\subsection{A Simple Example} \label{sec:ExampleBoundedRat}
\textit{Ingredients.} Consider the probabilistic model shown in Fig \ref{fig:graphmodel} with the joint distribution $p_0(\mathbf{X},\mathbf{S},A)$ that is specified by the factors in the decomposition \eqref{eq:exampleJoint}. Here, $S$ and $X$ denote the current environmental state and the corresponding observation, and $A$ denotes the action that must be determined in order to drive the system into a new state $S'$ with observation $X'$. The decision-making problem is specified by assuming that we have given a utility function $U$ over future observations $X'$ which the decision-maker seeks to maximize by selecting an action $A$, while only having access to the current observation $X$. 
This means that the decision-maker has control over the distribution $p(A|X)$, which replaces the prior $p_0(A)$ in the factorization \eqref{eq:exampleJoint} of the prior model $p_0(\mathbf{X},\mathbf{S},A)$ to determine the factorization of the posterior model $p(\mathbf{X},\mathbf{S},A)$ in terms of the fixed components in $p_0$ (cf. Fig \ref{fig:boundedRatExample}) as
\begin{equation} \label{eq:BRposterior}
p(\mathbf{X},\mathbf{S},A) =  \underbrace{p_0(X'|S') \, p_0(X|S)\, p_0(S'|S,A) \, p_0(S)}_{p_0(\mathbf{X},\mathbf{S}|A)}  \, p(A|X)\, .
\end{equation}

\noindent \textit{Free energy from constraints.} Further assuming that the decision-maker is subject to an information processing constraint $D_{\mathrm{KL}}(p\|p_0)\,{\leq}\, C_0$, for some non-negative bound $C_0$, results in the unconstrained optimization problem $\max_p F(p)$ with free energy given by \eqref{eq:broundedrational-general}, where the trade-off parameter $\beta$ is tuned to comply with the bound $C_0$. Since the action distribution $p(A|X)$ is the only distribution in the posterior model \eqref{eq:BRposterior} that changes during decision-making, i.e., during the transformation from prior to posterior, the total free energy simplifies to 

\vspace{-10pt}
\begin{align*}
F(p) & = \langle U \rangle_{p(\mathbf{X},\mathbf{S},A)} - \tfrac{1}{\beta} \DKL(p(\mathbf{X},\mathbf{S},A)\|p_0(\mathbf{X},\mathbf{S},A)) \\
     & = \langle V(X,A)\rangle_{p(A|X)p(X)} - \tfrac{1}{\beta}  \big\langle \DKL(p(A|X)\|p_0(A)) \big\rangle \\
     & = \langle F_A(p(A|X))\rangle_{p(X)} \,,
\end{align*}
where we have written $p_0(x|s)p_0(s) = p(s|x) p(x)$ using Bayes rule \eqref{eq:bayes-posterior}, and

\vspace{-10pt}
\begin{align*}
V(X,A) & \coloneqq \sum\nolimits_{s,s',x'} p(s|X)\, p_0(s'|s,A)\, p_0(x'|s')\, U(x') \, ,\\
F_A(p(A|X)) & \coloneqq \langle V(X,A)\rangle_{p(A|X)} - \tfrac{1}{\beta} D_\mathrm{KL}(p(A|X)\|p_0(A)) \, .
\end{align*}
Note that, here the expectation with respect to $p(X)$ does not affect the optimization with respect to $p(A|X)$ since it can be performed pointwise for each particular realization $x$ of $X$. In fact, we would have obtained the same result when conditioning on an arbitrary value $X\eqq x$ from the outset. However, in general, optimal information processing strategies may depend on the entire distribution $p(X)$ and can therefore not be obtained from only considering single observations $x$, for example when also optimizing with respect to the prior $p_0(A)$, see e.g., \citep{Genewein2015}.

\smallskip
\noindent \textbf{Free energy maximization.} The optimal action distribution $p^\ast(A|X)$ maximizing $F_A$ is a Boltzmann distribution \eqref{def:thermodyn:boltzmann} with ``energy'' $V(X,A)$ and prior $p_0(A)$, 
\begin{equation}
p^\ast(A|X) = \frac{1}{\mathcal Z(X)}\, p_0(A) \, e^{\beta V(X,A)} \, , \label{eq:BRsolution}
\end{equation}
where $\mathcal Z(X)\coloneqq \sum_a p_0(a) e^{\beta V(X,a)}$. Note that in order to evaluate the utility $V$, it is required to determine the Bayes' posterior $p(S|X)$. This shows how in a utility-based approach, the need to perform Bayesian inference results directly from the assumption about which variables are observed and which are not.

\subsection{Critical points} \label{sec:boundedrat:critical}

The main idea of free energy in the context of information processing with limited resources is that any computation can be thought of abstractly as a transformation from a distribution $p_0$ of prior knowledge to a posterior distribution $p$ that encapsulates an advanced state of knowledge resulting from deliberation. The progress that is made through such a transformation is quantitatively captured by two measures: the expected utility $\langle U\rangle_p$ that quantifies the quality of $p$ and $C(p,p_0)$ that measures the cost of uncertainty reduction from $p_0$ to $p$. Clearly, the critical point of this framework is the choice of the cost function $C$. In particular, we could ask whether there is some kind of universal cost function that is applicable to any perception-action process or whether there are only problem-specific instantiations. Of course, having a universal measure that allows applying the same concepts to extremely diverse systems is both a boon and a bane, because the practical insights it may provide for any concrete instance could be very limited. This is the root of a number of critical issues:

\begin{enumerate}[wide,labelindent=0pt]
\item[$(i)$] \textit{What is the cost $C$?} 
An important restriction of all deliberation costs of the form $C(p,p_0)$ is that they only depend on the initial and final distributions and ignore the process of how to get from $p_0$ to $p$. 
When varying a single resource (e.g., processing time) we can use $C(p,p_0)$ as a process-independent \emph{proxy} for the resource. However, if there are multiple resources involved (e.g., processing time, memory, and power consumption), a single cost cannot tell us how these resources are weighted optimally without making further process-dependent assumptions. 
In general, the theory makes no suggestions whatsoever about mechanical processes that could implement resource-optimal strategies, it only serves as a baseline for comparison.
Finally, simply requiring the measure to be monotonic in the uncertainty reduction, does not uniquely determine the form of $C$, as there have been multiple proposals of uncertainty measures in the literature (see e.g.~\citep{Csiszar2008}), where relative entropy is just one possibility. However, relative entropy is distinguished from all other uncertainty measures in its additivity property, that for example allows to express optimal probabilistic updates from $p_0$ to $p$ in terms of additions or subtractions of utilities, such as log-likelihoods for evidence accumulation in Bayesian inference.

\item[$(ii)$] \textit{What is the utility?}
When systems are engineered, utilities are usually assumed to be given such that desired behavior is specified by utility maximization. However, when we observe perception-action systems, it is often not so clear what the utility should be, or in fact, whether there even exists a utility that captures the observed behavior in terms of utility maximization. This question of the identifiability of a utility function is studied extensively in the economic sciences, where the basic idea is that systems reveal their preferences through their actual choices and that these preferences have to satisfy certain consistency axioms in order to guarantee the existence of a utility function.  In practice, to guarantee unique identifiability these axioms are usually rather strong, for example ignoring the effects of history and context when choosing between different items, or ignoring the possibility that there might be multiple objectives. When not making these strong assumptions, utility becomes a rather generic concept, like the concept of probability, and additional assumptions like soft-maximization are necessary to translate from utilities to choice probabilities.

\item[$(iii)$] \textit{The problem of infinite regress.}
One of the main conceptual issues with the interpretation of $C$ as a deliberation cost is that the original utility optimization problem is simply replaced by another optimization problem that may even be more difficult to solve. This novel optimization problem might again require resources to be solved and could therefore be described by a higher-level deliberation cost, thus leading to an infinite regress. In fact, any decision-making model that assumes that decision-makers reason about processing resources are affected by this problem \citep{Russel1995,Gigerenzer2001}. A possible way out is to consider the utility-information trade-off simply an \textit{as if} description, since perception-action systems that are subject to a utility-information trade-off do not necessarily have to reason or know about their deliberation costs.
It is straightforward, for example, to design processes that probabilistically optimize a given utility with no explicit notion of free energy, but for an outside observer the resulting choice distribution looks like an optimal free energy trade-off \citep{Ortega2014b}. 

\end{enumerate}
 
In summary, the free energy trade-off between utility and information primarily serves as a normative model for optimal probability assignments in information-processing nodes or networks. Like other Bayesian approaches, it can also serve as a guide for constructing and interpreting systems, although it is in general not a mechanistic model of behavior. In that respect it shares the fate of its cousins in thermodynamics and coding theory \citep{Shannon1948} in that they provide theoretical bounds on optimality but devise no  mechanism for processes to achieve these bounds.

\medskip

\section{Variational free energy in Active Inference} \label{sec:freeenergyprinciple}

\subsection{The basic idea}
Variational free energy is the main ingredient used in the \emph{Free Energy Principle} for biological systems in the neuroscience literature \citep{Friston2005,Friston2010,Friston2015,Friston2006}, which has been considered as ``arguably the most ambitious theory of the brain available today'' \citep{Gershman2019}. 
Since variational free energy in itself is just a mathematical construct to measure the dissimilarity between distributions and functions---see Section~\ref{sec:freeenergy}---, the biological content of the Free Energy Principle must come from somewhere else. The basic biological phenomenon that the Free Energy Principle purports to explain is \textit{homeostasis}, the ability to actively maintain certain relevant variables (e.g., blood sugar) within a preferred range. Usually, homeostasis is applied as an explanatory principle in physiology whereby the actual value of a variable is compared to a target value and corrections to deviation errors are made through a feedback loop. However, homeostasis has also been proposed as an explanatory principle for complex behavior in the cybernetic literature \citep{Wiener1948,Ashby1960,Powers1973,Cisek1999}---for example, maintaining blood sugar may entail complex feedback loops of learning to hunt, to trade and to buy food. Crucially, being able to exploit the environment in order to attain favorable sensory states, requires implicit or explicit knowledge of the environment that could either be pre-programmed (e.g., insect locomotion) or learnt (e.g., playing the piano). 
    
The Free Energy Principle was originally suggested as a theory of cortical responses \citep{Friston2005} by promoting the free energy formulation of predictive coding that was introduced by Dayan and Hinton with the Helmholtz machine \citep{Hinton1995}. It found its most recent incarnation in what is known as \textit{Active Inference} that attempts to extend variational Bayesian inference to the problem of action selection. Here, the target value of homeostasis is expressed through a probability distribution $p_\mathrm{des}$ under which desired sensory states have a high probability. The required knowledge about the environment is expressed through a generative model $p_0$ that relates observations, hidden causes, and actions. As the generative model allows to make predictions about future states and observations, it enables to choose actions in such a way that the predicted consequences conform to the desired distribution. In Active Inference, this is achieved by merging the generative and the desired distributions, $p_0$ and $p_\mathrm{des}$, into a single reference function $\phi$ to which trial distributions $q$ over the unknown variables are fitted by minimizing the variational free energy $F(q\|\phi)$. This free energy minimization is analogous to variational Bayesian inference, where the reference is always given by a joint distribution evaluated at observed quantities (cf.~Section \ref{sec:variationalinference}). In the resulting homeostatic process, the trial distributions $q$ play the role of internal variables that are manipulated in order to achieve desired sensory consequences that are not directly controllable. Minimizing variational free energy by the alternating variation of trial distributions over actions $q_\text{Actions}$ and trial distributions over hidden states $q_\text{States}$, 

\vspace{-10pt}
\begin{align}
\underbrace{\min_{q_\text{Actions}} F(q\|\phi)}_{\textrm{Action}}\quad\textrm{and}  \quad \underbrace{\min_{q_\text{States}} F(q\|\phi)}_{\textrm{Perception}} ,
\end{align}
is then equated with processes of action and perception. 


\begin{figure*}\includegraphics[width=1\textwidth]{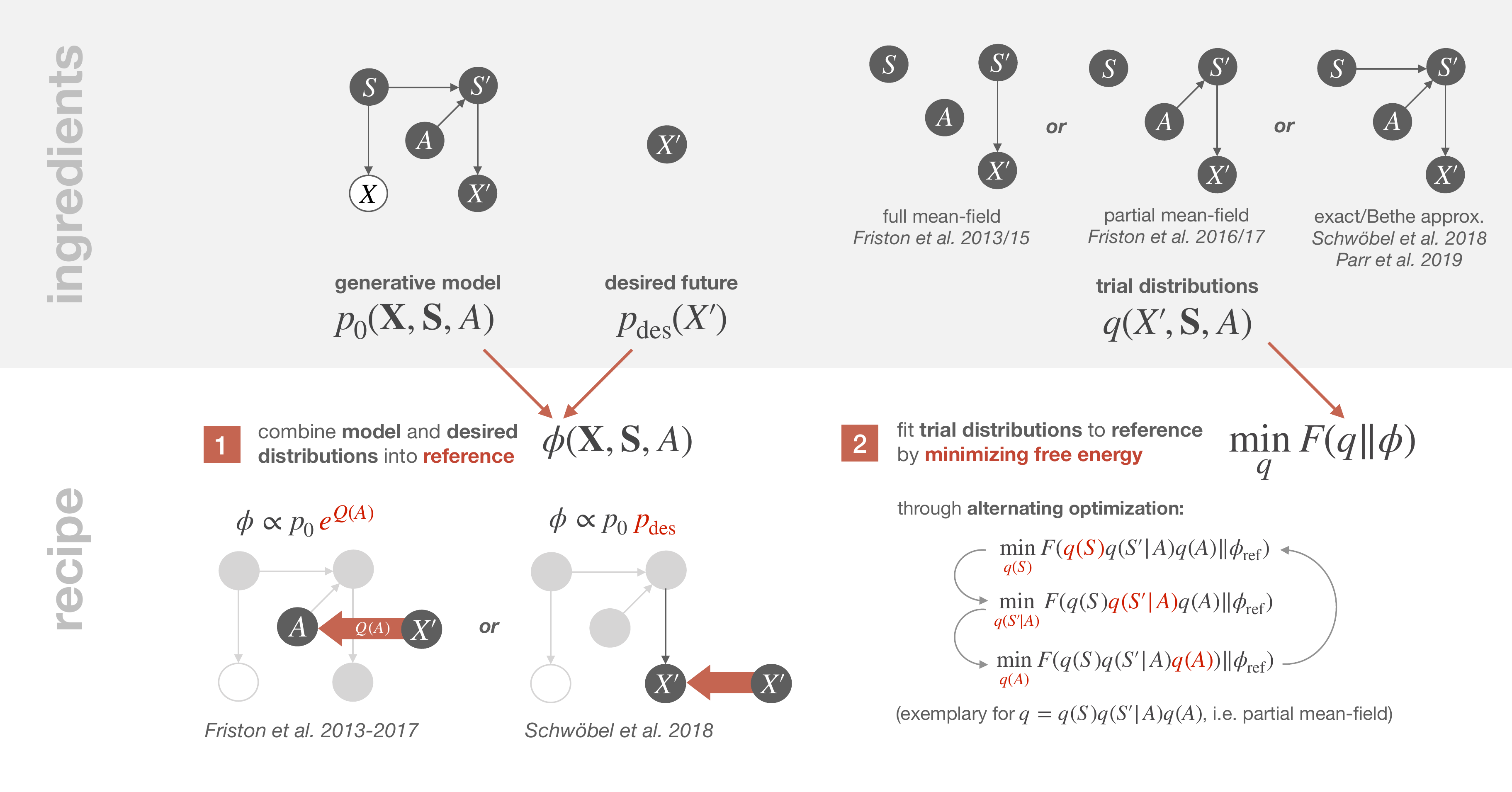}

\vspace{-5pt}
\caption{Overview of the Active Inference recipe, applied to our example from Fig \ref{fig:graphmodel}.}
\label{fig:activeInferenceExample}
\end{figure*}

In a nutshell, the central tenet of the Free Energy Principle states that organisms maintain homeostasis through minimization of variational free energy between a trial distribution $q$ and a reference function $\phi$ by acting and perceiving.
Sometimes the even stronger statement is made that minimizing variational free energy is \emph{mandatory} for homeostatic systems \citep{Friston2013L,Corcoran2018}.

\subsection{A Simple Example}\label{sec:ActiveInferenceExample}

\textit{Ingredients.} Applying the Active Inference recipe (cf.~Fig \ref{fig:activeInferenceExample}) to our running example from Fig~\ref{fig:graphmodel} with current and future states $S, S'$, current and future observations $X, X'$, and action $A$, we need a generative model $p_0$, a desired distribution $p_\mathrm{des}$, and trial distributions $q$. The generative model $p_0(\mathbf{X},\mathbf{S},A)$ is specified by the factors in the decomposition \eqref{eq:exampleJoint}, the desired distribution $p_\mathrm{des}(X')$ is a given fixed probability distribution over future sensory states $X'$, and the trial distributions $q$ are probability distributions over all unknown variables, $S,S',X'$, and $A$. 

In most treatments of Active Inference in the literature, the trial distributions $q$ are simplified, either by a full mean-field approximation over states and actions \citep{Friston2013,Friston2015}, by a partial mean-field approximation where the dependency on actions is kept but the states are treated independently of each other \citep{Friston2016,Friston2017b}, or more recently \citep{Schwoebel2018,Parr2019} by the so-called Bethe approximation \citep{Yedidia2001,Heskes2002}, where subsequent states are allowed to interact. In the partial mean-field assumption of \citep{Friston2016}, the trial distribution over $X'$ is fixed and given by $p_0(X'|S')$, while for $A$, $S$ and $S'$ the trial distributions are variable but restricted to be of the mean-field form for $S$ and $S'$,  
\begin{equation} \label{eq:activeinf-mean-field}
q(\mathbf{S},A) =  q(S)\, q(S'|A)\, q(A),
\end{equation}
i.e., the hidden states $S$ and $S'$ are assumed to be independent given $A$. While mean-field approximations can be good enough for simple perceptual inference, where a \textit{single} hidden cause might be responsible for a set of observations, they can be too strong simplifications for sequential decision-making problems where the next state $S'$ depends on the previous state $S$. In fact, as can be seen for example in \ref{S1_Sim}, mean-field assumptions may fail to show goal-directed behavior even for very simple tasks such as the navigation in a grid world. A less restrictive assumption would be a Bethe approximation, a special case of Kikuchi's cluster variation method \citep{Kikuchi1951}, which allows $S$ and $S'$ as well as $S'$ and $X'$ to be stochastically dependent---cf. Section C in Appendix \ref{S1_App}, where we derive the update equations under the Bethe assumption for the simple example of this section. In general, the Bethe approximation achieves exact marginals in tree-like models, such as the models that are considered in the Active Inference literature, because it results in update equations that are equivalent to Pearl's belief propagation algorithm \citep{Yedidia2001,Pearl1988}.

\noindent \textit{Reference function.} The reference $\phi$ is constructed by combining the two distributions $p_\mathrm{des}$ and $p_0$. To do so, there have been several proposals in the Active Inference literature, which fall into one of two categories: either a specific value function $Q$ is defined (containing $p_\mathrm{des}$), which is multiplied to the generative model using a soft-max function \citep{Friston2015,Friston2016,Friston2017b}, 
\begin{equation} \label{def:Qvaluereference}
\phi(X',\mathbf{S},A) \coloneqq p_0(X\eqq x,X',\mathbf{S}|A) \, \tfrac{1}{\mathcal{Z}} \,p_0(A) e^{Q(A)} \, ,
\end{equation}
or the desired distribution is multiplied directly to the generative model \citep{Schwoebel2018}, 
\begin{equation}
\phi(X',\mathbf{S},A) \coloneqq  p_\mathrm{des}(X') \, p_0(X\eqq x,X',\mathbf{S},A) \label{def:ControlAsInferenceReference}  .
\end{equation}

While the reference function in \eqref{def:ControlAsInferenceReference} is already completely specified, we still need to know how to determine the value function $Q$ in the case of \eqref{def:Qvaluereference}. For the partial mean-field assumption \eqref{eq:activeinf-mean-field} it is defined in the literature \citep{Friston2016,Friston2017b} as
\begin{equation} \label{def:Q-value}
Q(a) \coloneqq \langle U(X',S') \rangle_{q(X',S'|A\eqq a)} + H\big(q(X'|A\eqq a)\big),
\end{equation}
where $U(x',s') \coloneqq \log p_\mathrm{des}(x') + \log p_0(x'|s')$ favors both desirable and plausible future observations $x'$. While here desirability and plausibility is built into the value function $Q$ idiosyncratically, in utility-based approaches (cf.~Section \ref{sec:ExampleBoundedRat}) only desirability has to be put into the design of the utility function, because there the likelihood $p_0(X'|S')$ of future observations is automatically taken into account by the expected utility $V$ that is (soft-)maximized by \eqref{eq:BRsolution}. Moreover, since $Q$ can be rewritten as 
\[
Q(a) = -D_\mathrm{KL}\big(q(X'|A)\|p_\mathrm{des}(X')\big) - \big\langle H\big(p_0(X'|S')\big) \big\rangle_{q(S'|A)} \, ,
\]
the extra entropy term in \eqref{def:Q-value} has the effect of actions leading to consequences that more or less \emph{match} the desired distribution, while also explicitly punishing actions that lead to a high variability of observations (by requiring a low average entropy of $p_0(X'|S')$), rather than trying to produce the single most desired outcome---see the discussion at the end of Section \ref{sec:activeinference:critical}. Note also that the value function $Q$ depends (non-linearly) on the trial distribution $q(S'|A)$, because $q(X'|A)=\sum_{s'}p_0(X'|s')q(s'|A)$ is itself a function of $q(S'|A)$, which is problematic during free energy minimization (see $(ii)$ in Section \ref{sec:activeinference:critical}).

\smallskip
\noindent \textit{Free energy minimization.} Once the form of the trial distributions $q$---e.g.~by a partial mean-field assumption \eqref{eq:activeinf-mean-field} or a Bethe approximation (see \nameref{S1_App})---and the reference $\phi$ are defined, the variational free energy is simply determined by $F(q\|\phi)$. In the case of a mean-field assumption, the resulting free energy minimization problem is solved approximately by performing an alternating optimization scheme, in which the variational free energy is minimized separately with respect to each of the variable factors in a factorization of $q$, for example by alternating between $\min_{q(S)}F$, $\min_{q(S'|A)}F$, and $\min_{q(A)}F$ in the case of the partial mean-field assumption \eqref{eq:activeinf-mean-field}, where in each step the factors that are not optimized are kept fixed (cf. Fig \ref{fig:activeInferenceExample}). In \nameref{S1_App} we derive the update equations for the cases \eqref{def:Qvaluereference} and \eqref{def:ControlAsInferenceReference} under mean-field and Bethe approximations for the one-step example discussed in this section. Mean-field solutions for the general case of arbitrarily many timesteps together with their exact solutions can be found in \nameref{S1_Comp}, where we also highlight the theoretical differences between various proposed formulations of Active Inference. The effect of some of these differences can be seen in the grid world simulations in \nameref{S1_Sim}.

\subsection{Critical points} \label{sec:activeinference:critical}

The main idea behind Active Inference is to express the problem of action selection in a similar manner to the perceptual problem of Bayesian inference over hidden causes. In Bayesian inference, agents are equipped with likelihood models $p_0(X|Z)$ that determine the desirability of different hypotheses $Z$ under known data $X$. In Active Inference, agents are equipped with a given desired distribution $p_\mathrm{des}(X')$ over future outcomes that ultimately determines the desirability of actions $A$. An important difference that arises is that perceptual inference has to condition on past observations $X\eeqq x$, whereas naive inference over actions would have to condition on desired future outcomes $X'\eeqq x'$. 

For a single desired future observation $x'$, Bayesian inference could be applied in a straightforward way by simply conditioning the generative model $p_0$ on $X'\eqq x'$. Similarly, one could condition on a desired distribution $p_\mathrm{des}(X')$ using Jeffrey's conditioning rule \citep{Jeffrey1965}, resulting in $p(A|p_\mathrm{des}) \eeqq \sum_{x'} p(A|x')\, p_\mathrm{des}(x')$, which could be implemented by first sampling a goal $x'\,{\sim}\, p_\mathrm{des}(X')$ and then inferring $p(A|x')$ given the single desired observation $x'$. However, one of the problems with such a naive approach is that the choice of a goal is solely determined by its desirability, whereas its realizability for the decision-maker is not taken into account. This is because by conditioning on $p_\mathrm{des}$, the decision-maker effectively seeks to choose actions in order to \textit{reproduce} or \textit{match} the desired distribution.

\begin{figure*}{}
\includegraphics[width=1\textwidth]{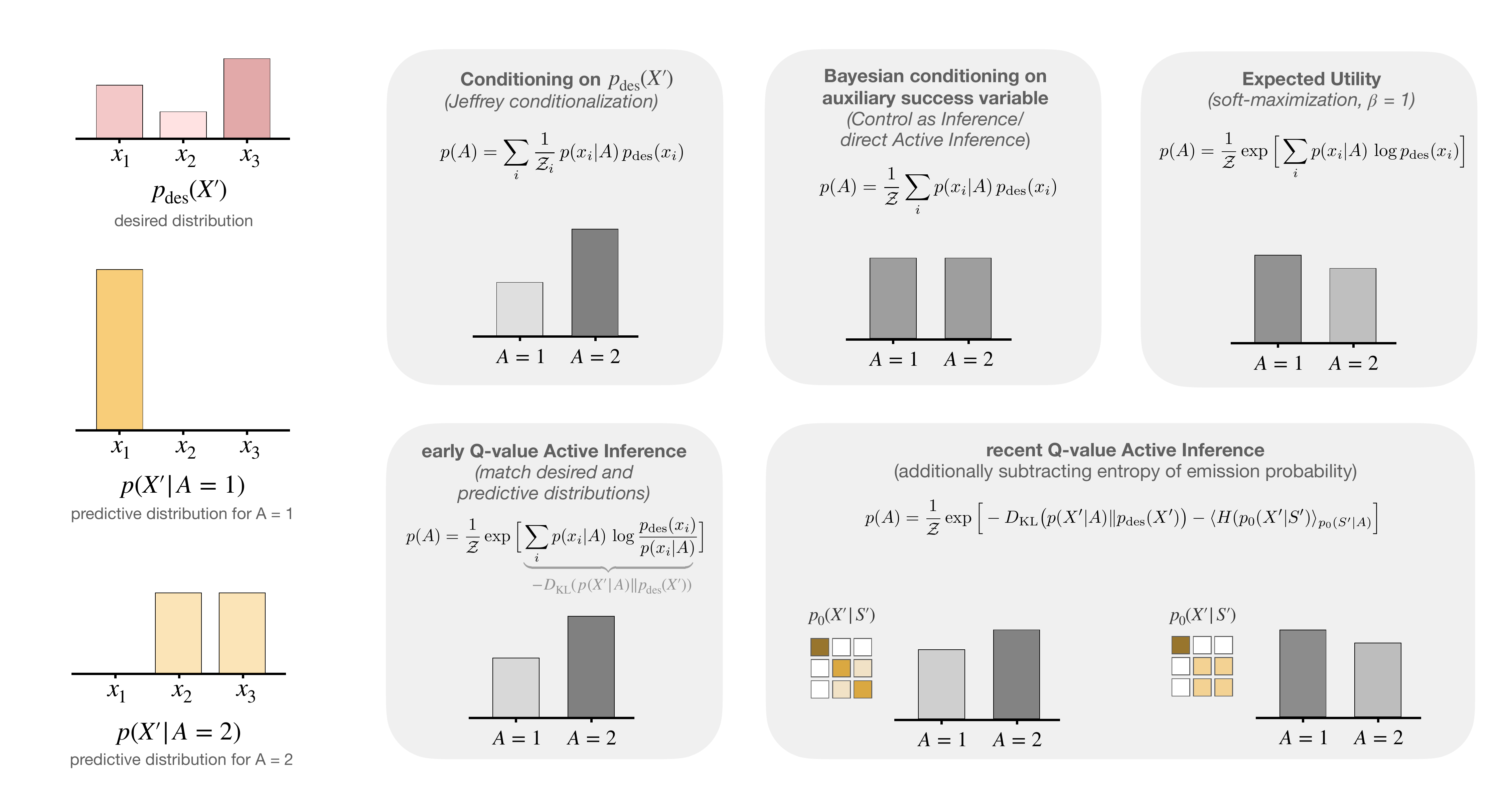}
\caption{Consequences of assuming a desired distribution $p_\mathrm{des}$ for action planning under purely inference-based methods, expected utility, and Active Inference, in the case of a simple example with two actions, one with a deterministic outcome and one with random outcomes. As can be seen from the displayed equations, conditioning on $p_\mathrm{des}$ (Jeffrey conditionalization) and conditioning on success (Control as Inference/direct Active Inference) only differ in the order of normalizing and taking the expectation over $X'$. While conditioning on $p_\mathrm{des}$ requires to first sample a target outcome from $p_\mathrm{des}$ before an action from $p(A|x')$ can be planned, conditioning on success directly weighs the desirability of an outcome $p_\mathrm{des}(x')$ by its realizability $p(x'|A)$. From this point of view, the expected utility approach is very similar to Control as Inference (which can also be seen in the grid world environment \nameref{S1_Sim}), since it also weighs the utility of an outcome with its realizability before soft-maximizing. It only differs in how it treats the desired distribution as an exponentiated utility, moving the utility values closer together so that option $A=1$ is slightly preferred. The early version \citep{Friston2013} of Active Inference is similar to Jeffrey conditioning, because decision-makers are also assumed to \textit{match} the desired distribution, by defining the value function $Q$ as a KL divergence between the predicted and desired distributions. In later versions of $Q$-value Active Inference \citep{Friston2015,Friston2016,Friston2017b}, the value function $Q$ is modified by an additional entropy term that explicitly punishes observations with high variability. Consequently, even when the effect of the action on future observations is kept the same, i.e., the predictive distribution \smash{$p(X'|A) = \sum_{s'}p_0(X'|s')p_0(s'|A)$} remains as depicted in the left-hand column, the preference over actions now changes completely depending on $p_0(X'|S')$---whereas in the other approaches, only the predictive distribution $p(X'|A)$ and $p_\mathrm{des}(X')$ influence planning. While there might be circumstances where this extra punishment of high outcome variability could be beneficial, it is  questionable from a normative point of view why anything else other than the predicted outcome probability $p(X'|A)$ should be considered for planning. See \nameref{S2_App} for details about the choices made in the example.}
\label{fig:exampleAppendix} 
\end{figure*}

To overcome this problem, \textit{Control as Inference} or \textit{Planning as Inference} approaches in the machine learning literature \citep{Toussaint2006,Todorov2008,Kappen2012,Levine2018,ODonoghue2020} do not directly condition on desired future observations but on future \textit{success} by introducing an auxiliary binary random variable $R$ such that $R=1$ encodes the occurence of desired outcomes. The auxiliary variable $R$ comes with a probability distribution $p_0(R|X',...)$ that determines how well the outcomes satisfy desirability criteria of the decision-maker, usually defined in terms of the reward or utility attached to certain outcomes---see the discussion in $(iii)$ below. The extra variable gives the necessary flexibility to infer successful actions by simply conditioning on $R\eqq 1$. The advantage of such an approach over direct Jeffrey conditionalization given a desired distribution over future observations can be seen in the grid world simulations in \nameref{S1_Sim}, especially the ability of choosing a desired outcome that is not only desirable but also achievable---see also Fig \ref{fig:exampleAppendix}.

Active Inference tries to overcome the same problem of reconciling realizability and desirability, but without explicitly introducing extra random variables and without explicitly conditioning on the future. Instead, the desired distribution is combined with the generative model to form a new reference function $\phi$ such that the posteriors $q^\ast$ resulting from the minimization of the free energy $F(q\|\phi)$ contain a baked-in tendency to reach the desired future encoded by $\phi$. This approach is the root of a number of critical issues with current formulations of Active Inference:

\begin{enumerate}[wide,labelindent=0pt]
\item[$(i)$] \textit{How to incorporate the desired distribution into the reference?}

\noindent Instead of using Bayesian conditioning directly in order to condition the generative model $p_0$ on the desired future, in Active Inference it is required that the reference $\phi$ contains the desired distribution in a way such that actions sampled from the resulting posterior model are more likely if they lead to the desired future. As can be seen already for the one-step case in \eqref{def:Qvaluereference} and \eqref{def:ControlAsInferenceReference}, the method of how to incorporate the desired distribution into the reference function is not unique and does not follow from first principles. There have been essentially two different proposals in the literature on Active Inference of how to combine the two distributions $p_\mathrm{des}$ and $p_0$ into $\phi$ (cf. Fig \ref{fig:activeInferenceExample}): Either a hand-crafted value function $Q$ is designed that specifically modifies the action probability of the generative model, or the probability over futures $X'$ under the generative model $p_0$ is modified by directly multiplying $p_\mathrm{des}$ to the likelihood $p_0(X'|S')$. We discuss both of these proposals in $(ii)$ and $(iii)$ below.

\item[$(ii)$] \textit{Proposal 1:} \textit{$Q$-value} Active Inference \citep{Friston2013,Friston2015,Friston2016,Friston2017b}
In the most popular formulation of Active Inference, the probability over actions in the reference $\phi$ is defined by $\frac{1}{\mathcal{Z}}\, p_0(A) e^{Q(A)}$, where the value function $Q$ (also called the ``expected free energy'')  depends non-linearly on the trial distributions $q$, as can be seen exemplarily in \eqref{def:Q-value} for the one-step case under the partial mean-field assumption of \cite{Friston2016,Friston2017b}, where $q(S'|A)$ enters $Q$ through $q(X'|A)=\sum_{s'}p_0(X'|s')q(s'|A)$. Note that, because of this non-linearity the alternating free energy minimization would have no closed-form solutions (cf.~\nameref{S1_App}). This means that both the trial distributions $q$ and the reference $\phi= \phi(q)$  will change when $q$ is varied during the minimization of the \emph{total} variational free energy $F(q\|\phi(q))$, as would be required when stipulating a single free energy functional for optimization. This highlights an important conceptual difference to variational Bayesian inference, where one assumes a \textit{fixed} reference $\phi$---resulting from the evaluation of a \textit{fixed} probabilistic model $p_0$ at known variables (see Section \ref{sec:variationalinference})---to which distributions $q$ are fitted by minimizing $F(q\|\phi)$. In contrast, when changing the reference $\phi(q)$ during the optimization process, it is no longer clear what is actually achieved by this minimization. As demonstrated by \nameref{S1_Sim}, this issue has immediate practical implications, as respecting or ignoring the extra $q$ dependency can result in very different behavior even in simple grid world simulations. \\
\indent In the Active Inference literature, however, the extra $q$-dependency of $Q$ is largely ignored. Instead of optimizing the full free energy $F(q\|\phi(q))$ with respect to state and action distributions, one alternatingly optimizes the free energy over states $F_A$ for each action $A$ and then the full free energy with respect to action distributions only, so that action and perception effectively optimize two different free energies. It is crucial to note, however, that unlike in variational Bayesian inference with fixed reference, this separation does not follow from the formalism of variational free energy, but is a design choice of the Active Inference framework that imposes this separation by force (see the Appendix \nameref{S4_App} for more details). This way, both separate optimizations can be considered as variational inference in each single update, even though when alternating them the reference $\phi$ still changes across the combined optimization process. This is in contrast to alternating optimization schemes in variational inference (e.g.,~in the Bayesian EM algorithm) where the reference $\phi$ does not change between optimization steps. Thus, there are two choices: Either Q-value Active Inference is regarded as some kind of approximation to variational inference under a single total free energy, or one has to give up the idea of a single free energy function that is optimized. Either way, the combined process of action and perception does not correspond to a single variational inference process.  \\
\indent Finally, another important practical issue with $Q$-value Active Inference models is that the definition of $Q$ relies on a mean-field approximation of the trial distributions $q$, under which hidden states are assumed to be stochastically independent. This simplification is too strong for sequential decision-making tasks, which renders the approach unfit for environments where the current state depends stochastically on previous states (see \nameref{S1_Sim} for a demonstration). 

\item[$(iii)$] \textit{Proposal 2:} \textit{direct} Active Inference \citep{Schwoebel2018}

\noindent When multiplying $p_\mathrm{des}$ to the generative model directly, as in \eqref{def:ControlAsInferenceReference}, then the resulting reference $\phi$ is no longer given by a joint distribution of observations, states, and actions (since in general $\sum_{x'}p_\mathrm{des}(x')p_0(x'|S') \not= 1$). Instead, this formulation of Active Inference turns out to be a special case of previous Control as Inference approaches in the machine learning literature \citep{Toussaint2006,Levine2018}, where one conditions on an auxiliary success variable $R$. In particular, for our running example from Fig \ref{fig:graphmodel} with a probabilistic model of the form \eqref{eq:exampleJoint}, Control as Inference defines 
\[
p_0(R=1|X',S',A) \coloneqq e^{r(X',S',A)} = 1 - p_0(R\eqq 0|X',S',A) \, ,
\]
where $r=r(X',S',A)$ denotes a general (negative) reward function determining desirability. The full joint of the new set of variables is then given by 
\begin{equation} \label{def:jointControlAsInference}
p_0(R,\mathbf{X},\mathbf{S},A) = p_0(R|X',S',A) \, p_0(\mathbf{X},\mathbf{S},A).
\end{equation}
Control as Inference then conditions actions on both, the history and future success ($R=1$). For our one-step example, this results in the Bayes' posterior
\begin{equation} \label{eq:exactControlAsInference}
p(A|X\eqq x, R\eqq 1) = \frac{1}{\mathcal{Z}} \sum_{x',s,s'} p_0(R=1|x',s',A)\ p_0(\mathbf{x},\mathbf{s},A) \, .
\end{equation}
It is straightforward to identify $p_\mathrm{des}(X')$ of Active Inference as a particular choice of a success probability $p_0(R\eqq 1|X')$, or equivalently, $\log p_\mathrm{des}(X')$ as a reward function $r=r(X')$, so that the joint distribution \eqref{def:jointControlAsInference} reduces to the reference function $\phi$ in \eqref{def:ControlAsInferenceReference}. Thus, the version of Active Inference in \citep{Schwoebel2018} is simply a variational formulation of Control as Inference that approximates exact posteriors of the form \eqref{eq:exactControlAsInference}, like other previous variational Bayes' approaches \citep{Toussaint2009,Ziebart2010,Levine2018}.

\end{enumerate}

In summary, the assumption of a desired distribution $p_\mathrm{des}$ over future outcomes has led to various attempts in the Active Inference literature of using probabilistic inference to determine profitable actions. Either an action distribution $\frac{1}{\mathcal{Z}}\,p_0(A)e^{Q(A)}$ is built into the reference function, which presupposes optimal behavior by designing a value function $Q$ that leads to desired consequences, or the outcome probability under the generative model $p_0$ is modified directly by multiplying $p_\mathrm{des}$ to $p_0$. The latter case is the variational version of Control as Inference, well-known in the machine learning literature \citep{Toussaint2006,Todorov2008,Toussaint2009,Ziebart2010,Kappen2012,Levine2018,ODonoghue2020}. Considering the issues of $Q$-value Active Inference discussed above, and the fact that Control as Inference does not rely on a desired distribution over outcomes, we could ask whether formulating preferences by assuming a desired distribution is well-advised. As can be seen from Fig \ref{fig:exampleAppendix}, the difference between purely inference-based methods, expected utility approaches, and Active Inference is mainly in how they treat the desired distribution. Should $p_\mathrm{des}$ be matched or is it good enough if actions are chosen that lead to a high desired outcome probability? While Control as Inference and utility-based models essentially take the latter approach, $Q$-value Active Inference answers this question by requiring that the desired distribution should be matched as long as the average entropy of $p_0(X'|S')$ is small.

\section{So What Does Free Energy Bring To the Table?}

\subsection{A Practical Tool}

It is unquestionable that the concept of free energy has seen many fruitful practical applications outside of physics in the statistical and machine learning literature. As has been discussed in Section \ref{sec:freeenergy}, these applications generally fall into one of two categories, the principle of maximum entropy, and a variational formulation of Bayesian inference. Here, the principle of maximum entropy is interpreted in a wider sense of optimizing a trade-off between uncertainty (entropy) and the expected value of some quantity of interest (energy), which in practice often appears in the form of regularized optimization problems (e.g., to prevent overfitting) or as a general inference method allowing to determine unbiased priors and posteriors (cf.~Section \ref{sec:maxent}). In the variational formulation of Bayes' rule, free energy plays the role of an error measure that allows to do approximate inference by constraining the space of distributions over which free energy is optimized, but can also inform the design of efficient iterative inference algorithms that result from an alternating optimization scheme where in each step the full variational free energy is optimized only partially, such as the Bayesian EM algorithm, belief propagation, and other message passing algorithms (cf.~Section \ref{sec:varFE}).

It is important to realize that, while the mathematical expressions of a free energy from constraints with ``energy'' $\mathcal E$ and trade-off parameter $\beta$ and a variational free energy with reference $\phi$ can formally be transformed into each other by $\phi = e^{-\beta \mathcal E}$, the two kinds of free energy are inherently distinct, both methodically and by their motivation. In the case of the free energy from constraints, we are given a constraint on some quantity $\mathcal{E}$ and we are trying to fulfil this constraint with minimum bias by selecting a distribution that trades off the two competing terms $\mathcal{E}$ and entropy. This trade-off also gives the reason for the existence of the Lagrange multiplier $\beta$ that has to be determined according to the constraint. In this sense the free energy from constraints is just a special case of the far more general Lagrangian method when applied to the optimization of expected values $\langle \mathcal E \rangle_p$ under entropy constraints (or the other way around). In contrast, variational free energy is simply a tool to represent the normalization of a reference function $\phi$ in terms of an optimization problem, and therefore does a priori not assume the existence of some quantity $\mathcal{E}$ that we may have observed in an experiment or that has any other constraints attached, nor does one explicitly consider entropy to be constrained or optimized. Therefore, even though starting from a (positive) reference function $\phi$ we can always invent the existence of some quantity $\mathcal{E}$ and some multiplier $\beta$ such that $\phi = e^{-\beta \mathcal E}$, this does not explain why these quantities should exist or why they should be mapped into each other in that particular way. The Lagrangian method, on the other hand, explains why for a given constraint on $\mathcal{E}$ we have a Lagrange multiplier $\beta$, how it is determined, and why the equilibrium distribution has the form $p^\ast = \frac{1}{\mathcal Z} e^{-\beta \mathcal E}$.

\subsection{Theories of Intelligent Agency}
These practical use-cases of free energy formulations have also influenced models of intelligent behavior. In the cognitive and behavioral sciences, intelligent agency has been modelled in a number of different frameworks, including logic-based symbolic models, connectionist models, statistical decision-making models, and dynamical systems approaches. Even though statistical thinking in a broader sense can in principle be applied to any of the other frameworks as well, statistical models of cognition in a more narrow sense have often focused on Bayesian models, where agents are equipped with probabilistic models of their environment allowing them to infer unknown variables in order to select actions that lead to desirable consequences \citep{Tenenbaum2001,Wolpert2006,Todorov2009}. Naturally, the inference of unknown variables in such models can be achieved by a plethora of methods including the two types of free energy approaches of maximum entropy and variational Bayes. However, both free energy formulations go one step further in that they attempt to extend both principles from the case of inference to the case of action selection: utility optimization with information constraints based on free energy from constraints and Active Inference based on variational free energy. 

While sharing similar mathematical concepts, both approaches differ in syntax and semantics. An apparent apple of discord is the concept of utility \citep{Gershman2012}. Utility optimization with information constraints requires the determination of a utility function, whereas Active Inference requires the determination of a reference function. In the economic literature, subjective utility functions that quantify the preferences of decision-makers are typically restrictive in order to ensure identifiability when certain consistency axioms are satisfied. In contrast, in Active Inference the reference function involves determining a desired distribution given by the preferred frequency of outcomes. However, these differences start to vanish when weakening the utility concept to something like log-probabilities, such that the utility framework becomes more similar to the concept of probability that is able to explain arbitrary behavior. Moreover, Active Inference has to solve the additional problem of marrying up the agent's probabilistic model with its desired distribution into a single reference function (cf.~Section \ref{sec:activeinference:critical}). The solution to this problem is not unique, in particular it lies outside the scope of variational Bayesian inference, but it is critical for the resulting behavior because it determines the exact solutions that are approximated by free energy minimization. In fact, as can be seen in simple simulations such as \ref{S1_Sim}, the various proposals for this merging that can be found in the Active Inference literature behave very differently.

Also, both approaches differ fundamentally in their motivation. The motivation of utility optimization with information constraints is to capture the trade-off between precision and uncertainty that underlies information processing. This trade-off takes the form of a free energy once an informational cost function has been chosen (cf.~Section \ref{sec:boundedrat:critical}). Note that Bayes' rule can be seen as the minimum of a free energy from constraints with log-likelihoods as utilities, even though this equivalence is not the primary motivation of this trade-off. In contrast, Active Inference is motivated from casting the problem of action selection itself as an inference process \citep{Friston2013}, as this allows to express both action and perception as the result of minimizing the same function, the variational free energy. However, there is no mystery in having such a single optimization function, because the underlying probabilistic model already contains both action and perception variables in a single functional format and the variational free energy is just a function of that model. Moreover, while approximate inference can be formulated on the basis of variational free energy, inference in general does not rely on this concept, in particular inference over actions can easily be done without free energy \citep{Dayan1997,Toussaint2006,Todorov2008,Kappen2012,Levine2018}. 

However, there are also plenty of similarities between the two free energy approaches. For example, the assumption of a soft-max action distribution in Active Inference is similar to the posterior solutions resulting from utility optimization with information constraints. Moreover, the assumption of a desired future \emph{distribution} relates to constrained computational resources, because the uncertainty constraint in a desired distribution over future states may not only be a consequence of environmental uncertainty, but could also originate from stochastic preferences of a satisficing decision-maker that accepts a wide range of outcomes. In fact, as we have seen in the discussion around Fig \ref{fig:exampleAppendix}, various methods for inference over actions differ in how they treat preferences given by a distribution over desired outcomes: Some of them try to match the predictive and desired distributions, while others simply seek to reach states whose outcomes have a high desired probability. In \ref{S1_Sim}, we provide a comparison of the discussed methods using grid world simulations, in order to see their resulting behavior also in a sequential decision-making task.

A remarkable resemblance among both approaches is the exclusive appearance of relative entropy to measure dissimilarity. In the Active Inference literature it is often claimed that every homeostatic system must minimize variational free energy \citep{Friston2013L}, which is simply an extension of relative entropy for non-normalized reference functions (cf.~Section \ref{sec:normalization}). In utility-based approaches, the relative entropy \eqref{eq:informationcost} is typically used to measure the amount of information processing, even though theoretically other cost functions would be conceivable \citep{Gottwald2019}. For a given homeostatic process, the KL divergence measures the dissimilarity between the current distribution and the limiting distribution and therefore is reduced while approximating the equilibrium. Similarly, in utility-based decision-making models, relative entropy measures the dissimilarity between the current posterior and the prior. In the Active Inference literature the stepwise minimization of variational free energy that goes along with KL minimization is often equated with the minimization of sensory \textit{surprise} (see \nameref{S3_App} for a more detailed explanation), an idea that stems from maximum likelihood algorithms, 
but that has been challenged as a general principle (see \citep{Biehl2020} and the response \citep{Friston2020}). Similarly, one could in principle rewrite free energy from constraints in terms of informational surprise, which would however simply be a rewording of the probabilistic concepts in log-space. The same kind of rewording is well-known between probabilistic inference and the minimum description length principle \citep{Gruenwald2007} that also operates in log-space, and thus reformulates the inference problem as a surprise minimization problem without adding any new features or properties.

\subsection{Biological Relevance}

So far we have seen how free energy is used as a technical instrument to solve inference problems and its corresponding appearance in different models of intelligent agency. Crucially, these kinds of models can be applied to any input-output system, be it a human that reacts to sensory stimuli, a cell that tries to maintain homeostasis, or a particle trapped by a physical potential. Given the existing literature that has widely applied the concept of free energy to biological systems, we may ask whether there are any specific biological implications of these models.

Considering free energy from constraints, the trade-off between utility and information processing costs provides a normative model of decision-making under resource constraints, that extends previous optimality models based on expected utility maximization and Bayesian inference. Analogous to rate-distortion curves in information theory, optimal solutions to decision-making problems are obtained that separate achievable from non-achievable regions in the information-utility plane (cf.~Fig \ref{fig:boundedrat}). The behavior of real decision-making systems under varying information constraints can be analyzed experimentally by comparing their performance with respect to the corresponding optimality curve. One can experimentally relate abstract information processing costs measured in bits to task-dependent resource costs like reaction or planning times \citep{Schach2018,Ortega2016a}. 
Moreover, the free energy trade-off can also be used to describe networks of agents, where each agent is limited in its ability, but the system as a whole has a higher information processing capacity---for example, neurons in a brain or humans in a group. In such systems different levels of abstraction arise depending on the different positions of decision-makers in the network \citep{LindigLeon2019, Genewein2015, Gottwald2018}. As we have discussed in Section \ref{sec:boundedrat:critical}, just like coding and rate-distortion theory, utility theory with information costs can only provide optimality bounds but does not specify any particular mechanism of how to achieve optimality. However, by including more and more constraints one can make a model more and more mechanistic and thereby gradually move from a normative to a more descriptive model, such as models that consider the communication channel capacity of neurons with a finite energy budget \citep{Gershman2018}. 

Considering variational free energy, there is a vast literature on biological applications mostly focusing on neural processing (e.g., predictive coding, dopamine) \citep{Schwartenbeck2015,Friston2017,Parr2019}, but there are also a number of applications aiming to explain behavior (e.g., human decision-making, hallucinations) \citep{Parr2018}. 
Similarly to utility-based models, Active Inference models can be studied in terms of \emph{as if} models, so that actual behavior can be compared to predicted behavior as long as suitable prior and likelihood models can be identified from the experiment. When applied to brain dynamics, the as if models are sometimes also given a mechanistic interpretation by relating iterative update equations that appear when minimizing variational free energy with dynamics in neuronal circuits. As discussed in Section \ref{sec:approxIterativeInf}, the update equations resulting for example from mean-field or Bethe approximations, can often be written in message passing form in the sense that the update for a given variable only has contributions that requires the current approximate posterior of neighbouring nodes in the probabilistic model. These contributions are interpreted as local messages passed between the nodes and might be related to brain signals \citep{Parr2019}. Other interpretations \citep{Friston2006,Friston2017b,Bogacz2017} obtain similar update equations by minimizing variational free energy directly through gradient descent, which can again be related to neural coding schemes like predictive coding. As these coding schemes have existed irrespective of free energy \citep{Rao1999,Aitchison2017}, especially since minimization of prediction errors is already seen in maximum likelihood estimation \citep{Rao1999}, the question remains whether there are any specific predictions of the Active Inference framework that cannot be explained with previous models (see \citep{Colombo2018,Hohwy2020} for recent discussions of this question).

\subsection{Conclusion}

Any theory about intelligent behavior has to answer three questions: \emph{Where am I?}, \emph{where do I want to go?}, and \emph{how do I get there?}, corresponding to the three problems of inference and perception, goals and preferences, and planning and execution. All three problems can be addressed either in the language of probabilities or utilities. Perceptual inference can either be considered as finding parameters that maximize probabilities or likelihood utilities. Goals and preferences can either be expressed by utilities over outcomes or by desired distributions. The third question can be answered by the two free energy approaches that either determine future utilities based on model predictions, or infer actions that lead to outcomes predicted to have high desired probability or match the desired distribution. In standard decision-making models actions are usually determined by a utility function that ranks different options, whereas perceptual inference is determined by a likelihood model that quantifies how probable certain observations are. In contrast, both free energy approaches have in common that they treat all types of information processing, from action planning to perception, as the same formal process of minimizing some form of free energy. But the crucial difference is not whether they use utilities or probabilities, but how predictions and goals are interwoven into action. 

This article started out by tracing back the seemingly mysterious connection between Helmholtz free energy from thermodynamics and Helmholtz' view of model-based information processing that led to the analysis-by-synthesis approach of perception, as exemplified in predictive coding schemes, and in particular to discuss the role of free energy in current models of intelligent behavior. The mystery starts to dissolve when we consider the two kinds of free energies discussed in this article, one based on the maximum entropy principle and the other based on variational free energy---a dissimilarity measure between distributions and (generally unnormalized) functions that extends the well-known KL divergence from information theory. The Helmholtz free energy is a particular example of an energy information trade-off that results from the maximum entropy principle \citep{Jaynes1957}. Analysis-by-synthesis is a particular application of inference to perception, where determining model parameters and hidden states can either be seen as a result of maximum entropy under observational constraints or of fitting parameter distributions to the model through variational free energy minimization. Thus, both notions of free energy can be formally related as entropy-regularized maximization of log-probabilities.

Conceptually, however, utility-based models with information constraints serve primarily as \emph{ultimate} explanations of behavior, this means they do not focus on mechanism, but on the goals of behavior and their realizability under ideal circumstances. They have the appeal of being relatively straightforward generalization of standard utility theory, but they rely on abstract concepts like utility and relative entropy that may not be so straightforwardly related to experimental settings. While these normative models have no immediate mechanistic interpretation, their relevance for mechanistic models may be analogous to the relevance of optimality bounds in Shannon's information theory for practical codes \citep{Shannon1948}. In contrast, Active Inference models of behavior often mix ultimate and \emph{proximate} arguments of explaining behavior \citep{Alcock1993,Tinbergen1963}, because they combine the normative aspect of optimizing variational free energy with the mechanistic interpretation of the particular form of approximate solutions to this optimization. While mean-field approaches of Active Inference may be particularly amenable to such mechanistic interpretations, they are often too simple to capture complex behavior. In contrast, the solutions of direct Active Inference resulting from a Bethe assumption are equivalent to previous Control as Inference approaches \citep{Toussaint2006,Todorov2008,Toussaint2009,Ziebart2010,Kappen2012,Levine2018,ODonoghue2020} that allow for Bayesian message passing formulations whose biological implementability can be debated irrespective of the existence of a free energy functional.

Finally, both kinds of free energy formulations of intelligent agency are so general and flexible in their ingredients that it might be more appropriate to consider them languages or tools to phrase and describe behavior rather than theories that explain behavior, in a sense similar to how statistics and probability theory are not biological or physical theories but simply provide a language in which we can phrase our biological and physical assumptions.

\section*{Funding}

This study was funded by the European Research Council (ERC-StG-2015-ERC Starting Grant, Project ID: 678082, ``BRISC: Bounded Rationality in Sensorimotor Coordination'').

\bigskip

{\footnotesize


\begin{thebibliography}{}

\bibitem[Aitchison and Lengyel, 2017]{Aitchison2017}
Aitchison, L. and Lengyel, M. (2017).
\newblock With or without you: predictive coding and bayesian inference in the
  brain.
\newblock {\em Current Opinion in Neurobiology}, 46:219--227.
\newblock Computational Neuroscience.

\bibitem[Alcock, 1993]{Alcock1993}
Alcock, J. (1993).
\newblock {\em Animal behavior: an evolutionary approach}.
\newblock Sinauer Associates.

\bibitem[Ashby, 1960]{Ashby1960}
Ashby, W. (1960).
\newblock {\em Design for a Brain: The Origin of Adaptive Behavior}.
\newblock Springer Netherlands.

\bibitem[Beal, 2003]{beal2003}
Beal, M.~J. (2003).
\newblock {\em Variational Algorithms for Approximate Bayesian Inference}.
\newblock PhD thesis, University of Cambridge, UK.

\bibitem[Bernoulli, 1713]{Bernoulli1713}
Bernoulli, J. (1713).
\newblock {\em Ars conjectandi}.
\newblock Basel, Thurneysen Brothers.

\bibitem[Bhui and Gershman, 2018]{Gershman2018}
Bhui, R. and Gershman, S.~J. (2018).
\newblock Decision by sampling implements efficient coding of psychoeconomic
  functions.
\newblock {\em Psychological Review}, 125(6):985--1001.

\bibitem[Biehl et~al., 2020]{Biehl2020}
Biehl, M., Pollock, F.~A., and Kanai, R. (2020).
\newblock A technical critique of the free energy principle as presented in
  "life as we know it" and related works.
\newblock {\em arXiv:2001.06408}.

\bibitem[Binz et~al., 2020]{Binz2020}
Binz, M., Gershman, S.~J., Schulz, E., and Endres, D. (2020).
\newblock Heuristics from bounded meta-learned inference.

\bibitem[Bogacz, 2017]{Bogacz2017}
Bogacz, R. (2017).
\newblock A tutorial on the free-energy framework for modelling perception and
  learning.
\newblock {\em Journal of Mathematical Psychology}, 76:198--211.
\newblock Model-based Cognitive Neuroscience.

\bibitem[Boutilier et~al., 1999]{Boutilier1999}
Boutilier, C., Dean, T., and Hanks, S. (1999).
\newblock Decision-theoretic planning: Structural assumptions and computational
  leverage.
\newblock {\em J. Artif. Int. Res.}, 11(1):1--94.

\bibitem[Cisek, 1999]{Cisek1999}
Cisek, P. (1999).
\newblock Beyond the computer metaphor: behaviour as interaction.
\newblock {\em Journal of Consciousness Studies}, 6(11-12):125--142.

\bibitem[Clark, 2013]{Clark2013}
Clark, A. (2013).
\newblock Whatever next? predictive brains, situated agents, and the future of
  cognitive science.
\newblock {\em Behavioral and Brain Sciences}, 36(3):181--204.

\bibitem[Colombo and Wright, 2018]{Colombo2018}
Colombo, M. and Wright, C. (2018).
\newblock First principles in the life sciences: the free-energy principle,
  organicism, and mechanism.
\newblock {\em Synthese}.

\bibitem[Corcoran and Hohwy, 2018]{Corcoran2018}
Corcoran, A.~W. and Hohwy, J. (2018).
\newblock {\em Allostasis, interoception, and the free energy principle:
  Feeling our way forward}.
\newblock Oxford University Press.

\bibitem[Csiszár, 2008]{Csiszar2008}
Csiszár, I. (2008).
\newblock Axiomatic characterizations of information measures.
\newblock {\em Entropy}, 10(3):261--273.

\bibitem[Csiszár and Tusnády, 1984]{Csiszar1984}
Csiszár, I. and Tusnády, G. (1984).
\newblock Information geometry and alternating minimization procedures.
\newblock {\em Statistics and Decisions, Supplement Issue}, 1:205--237.

\bibitem[Dayan and Hinton, 1997]{Dayan1997}
Dayan, P. and Hinton, G.~E. (1997).
\newblock Using expectation-maximization for reinforcement learning.
\newblock {\em Neural Computation}, 9(2):271--278.

\bibitem[Dayan et~al., 1995]{Hinton1995}
Dayan, P., Hinton, G.~E., Neal, R.~M., and Zemel, R.~S. (1995).
\newblock The helmholtz machine.
\newblock {\em Neural Comput.}, 7(5):889--904.

\bibitem[de~Laplace, 1812]{Laplace1812}
de~Laplace, P.~S. (1812).
\newblock {\em Théorie analytique des probabilités}.
\newblock Ve. Courcier, Paris.

\bibitem[Dempster et~al., 1977]{Dempster1977}
Dempster, A.~P., Laird, N.~M., and Rubin, D.~B. (1977).
\newblock Maximum likelihood from incomplete data via the em algorithm.
\newblock {\em Journal of the Royal Statistical Society. Series B
  (Methodological)}, 39(1):1--38.

\bibitem[Doya, 2007]{Doya2007}
Doya, K. (2007).
\newblock {\em Bayesian Brain: Probabilistic Approaches to Neural Coding}.
\newblock MIT Press, Cambridge, Mass.

\bibitem[Ergin and Sarver, 2010]{Ergin2010}
Ergin, H. and Sarver, T. (2010).
\newblock A unique costly contemplation representation.
\newblock {\em Econometrica}, 78(4):1285--1339.

\bibitem[Feynman et~al., 1996]{Feynman1996}
Feynman, R., Hey, A., and Allen, R. (1996).
\newblock {\em Feynman Lectures on Computation}.
\newblock Advanced book program. Addison-Wesley.

\bibitem[Flanagan et~al., 2003]{Flanagan2003}
Flanagan, J.~R., Vetter, P., Johansson, R.~S., and Wolpert, D.~M. (2003).
\newblock Prediction precedes control in motor learning.
\newblock {\em Current Biology}, 13(2):146--150.

\bibitem[Fox et~al., 2016]{Tishby2016}
Fox, R., Pakman, A., and Tishby, N. (2016).
\newblock Taming the noise in reinforcement learning via soft updates.
\newblock In {\em Proceedings of the Thirty-Second Conference on Uncertainty in
  Artificial Intelligence}, UAI'16, pages 202--211, Arlington, Virginia, United
  States. AUAI Press.

\bibitem[Friston, 2013]{Friston2013L}
Friston, K. (2013).
\newblock Life as we know it.
\newblock {\em Journal of The Royal Society Interface}, 10(86):20130475.

\bibitem[Friston, 2018]{Friston2018}
Friston, K. (2018).
\newblock Does predictive coding have a future?
\newblock {\em Nature Neuroscience}, 21(8):1019--1021.

\bibitem[Friston et~al., 2020]{Friston2020}
Friston, K., Costa, L.~D., and Parr, T. (2020).
\newblock Some interesting observations on the free energy principle.
\newblock {\em arXiv:2002.04501}.

\bibitem[Friston et~al., 2016]{Friston2016}
Friston, K., FitzGerald, T., Rigoli, F., Schwartenbeck, P., O'Doherty, J., and
  Pezzulo, G. (2016).
\newblock Active inference and learning.
\newblock {\em Neuroscience {\&} Biobehavioral Reviews}, 68:862--879.

\bibitem[Friston et~al., 2013]{Friston2013}
Friston, K., Schwartenbeck, P., Fitzgerald, T., Moutoussis, M., Behrens, T.,
  and Dolan, R. (2013).
\newblock The anatomy of choice: active inference and agency.
\newblock {\em Frontiers in Human Neuroscience}, 7:598.

\bibitem[Friston, 2005]{Friston2005}
Friston, K.~J. (2005).
\newblock A theory of cortical responses.
\newblock {\em Philosophical Transactions of the Royal Society B: Biological
  Sciences}, 360(1456):815--836.

\bibitem[Friston, 2010]{Friston2010}
Friston, K.~J. (2010).
\newblock The free-energy principle: a unified brain theory?
\newblock {\em Nature Reviews Neuroscience}, 11:127--138.

\bibitem[Friston et~al., 2017a]{Friston2017b}
Friston, K.~J., FitzGerald, T. H.~B., Rigoli, F., Schwartenbeck, P., and
  Pezzulo, G. (2017a).
\newblock Active inference: A process theory.
\newblock {\em Neural Computation}, 29:1--49.

\bibitem[Friston et~al., 2006]{Friston2006}
Friston, K.~J., Kilner, J., and Harrison, L.~M. (2006).
\newblock A free energy principle for the brain.
\newblock {\em Journal of Physiology-Paris}, 100:70--87.

\bibitem[Friston et~al., 2017b]{Friston2017}
Friston, K.~J., Parr, T., and de~Vries, B. (2017b).
\newblock The graphical brain: Belief propagation and active inference.
\newblock {\em Network Neuroscience}, 1(4):381--414.

\bibitem[Friston et~al., 2015]{Friston2015}
Friston, K.~J., Rigoli, F., Ognibene, D., Mathys, C., Fitzgerald, T., and
  Pezzulo, G. (2015).
\newblock Active inference and epistemic value.
\newblock {\em Cognitive Neuroscience}, 6(4):187--214.

\bibitem[Friston et~al., 2012]{Friston2012}
Friston, K.~J., Shiner, T., FitzGerald, T., Galea, J.~M., Adams, R., Brown, H.,
  Dolan, R.~J., Moran, R., Stephan, K.~E., and Bestmann, S. (2012).
\newblock Dopamine, affordance and active inference.
\newblock {\em PLoS Computational Biology}, 8(1):e1002327.

\bibitem[Garner, 1962]{Garner1962}
Garner, W.~R. (1962).
\newblock {\em Uncertainty and structure as psychological concepts}.
\newblock Wiley.

\bibitem[Genewein et~al., 2015]{Genewein2015}
Genewein, T., Leibfried, F., Grau-Moya, J., and Braun, D.~A. (2015).
\newblock Bounded rationality, abstraction, and hierarchical decision-making:
  An information-theoretic optimality principle.
\newblock {\em Frontiers in Robotics and AI}, 2.

\bibitem[Gershman, 2019]{Gershman2019}
Gershman, S.~J. (2019).
\newblock What does the free energy principle tell us about the brain.
\newblock {\em Neurons, Behavior, Data Analysis, and Theory}.

\bibitem[Gershman and Daw, 2012]{Gershman2012}
Gershman, S.~J. and Daw, N.~D. (2012).
\newblock Perception, action and utility: The tangled skein.
\newblock In {\em Principles of Brain Dynamics}. MIT Press.

\bibitem[Gigerenzer and Selten, 2001]{Gigerenzer2001}
Gigerenzer, G. and Selten, R. (2001).
\newblock {\em Bounded Rationality: The Adaptive Toolbox}.
\newblock MIT Press: Cambridge, MA, USA.

\bibitem[Gottwald and Braun, 2019a]{Gottwald2019}
Gottwald, S. and Braun, D.~A. (2019a).
\newblock Bounded rational decision-making from elementary computations that
  reduce uncertainty.
\newblock {\em Entropy}, 21(4).

\bibitem[Gottwald and Braun, 2019b]{Gottwald2018}
Gottwald, S. and Braun, D.~A. (2019b).
\newblock Systems of bounded rational agents with information-theoretic
  constraints.
\newblock {\em Neural Computation}, 31(2):440--476.

\bibitem[Grau-Moya et~al., 2016]{GrauMoya2016}
Grau-Moya, J., Leibfried, F., Genewein, T., and Braun, D.~A. (2016).
\newblock Planning with information-processing constraints and model
  uncertainty in markov decision processes.
\newblock In {\em Machine Learning and Knowledge Discovery in Databases}, pages
  475--491. Springer International Publishing.

\bibitem[Grünwald, 2007]{Gruenwald2007}
Grünwald, P. (2007).
\newblock {\em The Minimum Description Length Principle}.
\newblock MIT Press, Cambridge, Mass.

\bibitem[Haarnoja et~al., 2017]{Levine2017}
Haarnoja, T., Tang, H., Abbeel, P., and Levine, S. (2017).
\newblock Reinforcement learning with deep energy-based policies.
\newblock In {\em ICML}.

\bibitem[Hansen and Sargent, 2008]{Hansen2008}
Hansen, L.~P. and Sargent, T.~J. (2008).
\newblock {\em Robustness}.
\newblock Princeton University Press.

\bibitem[Harsha et~al., 2010]{Harsha2009}
Harsha, P., Jain, R., McAllester, D., and Radhakrishnan, J. (2010).
\newblock The communication complexity of correlation.
\newblock {\em {IEEE} Transactions on Information Theory}, 56(1):438--449.

\bibitem[Hathaway, 1986]{Hathaway1986}
Hathaway, R.~J. (1986).
\newblock Another interpretation of the em algorithm for mixture distributions.
\newblock {\em Statistics {\&} Probability Letters}, 4(2):53--56.

\bibitem[Heskes, 2003]{Heskes2002}
Heskes, T. (2003).
\newblock Stable fixed points of loopy belief propagation are local minima of
  the bethe free energy.
\newblock In Becker, S., Thrun, S., and Obermayer, K., editors, {\em Advances
  in Neural Information Processing Systems 15}, pages 359--366. MIT Press.

\bibitem[Hinton and van Camp, 1993]{Hinton1993}
Hinton, G.~E. and van Camp, D. (1993).
\newblock Keeping the neural networks simple by minimizing the description
  length of the weights.
\newblock In {\em Proceedings of the Sixth Annual Conference on Computational
  Learning Theory}, COLT '93, pages 5--13, New York, NY, USA. ACM.

\bibitem[Ho et~al., 2020]{Griffiths2020}
Ho, M.~K., Abel, D., Cohen, J.~D., Littman, M.~L., and Griffiths, T.~L. (2020).
\newblock The efficiency of human cognition reflects planned information
  processing.
\newblock {\em Proceedings of the 34th AAAI Conference on Artificial
  Intelligence}.

\bibitem[Hohwy, 2020]{Hohwy2020}
Hohwy, J. (2020).
\newblock Self-supervision, normativity and the free energy principle.
\newblock {\em Synthese}.

\bibitem[Jaynes, 1957]{Jaynes1957}
Jaynes, E.~T. (1957).
\newblock Information theory and statistical mechanics.
\newblock {\em Phys. Rev.}, 106:620--630.

\bibitem[Jaynes, 2003]{Jaynes2003}
Jaynes, E.~T. (2003).
\newblock {\em Probability Theory}.
\newblock Cambridge University Press.

\bibitem[Jeffrey, 1965]{Jeffrey1965}
Jeffrey, R.~C. (1965).
\newblock {\em The Logic of Decision}.
\newblock University of Chicago Press, 1 edition.

\bibitem[Kahneman, 2002]{Kahneman2002}
Kahneman, D. (2002).
\newblock Maps of bounded rationality: A perspective on intuitive judgement.
\newblock In Frangsmyr, T., editor, {\em Nobel prizes, presentations,
  biographies, {\&} lectures}, pages 416--499. Almqvist {\&} Wiksell,
  Stockholm, Sweden.

\bibitem[Kappen et~al., 2012]{Kappen2012}
Kappen, H.~J., Gómez, V., and Opper, M. (2012).
\newblock Optimal control as a graphical model inference problem.
\newblock {\em Machine Learning}, 87(2):159--182.

\bibitem[Kawato, 1999]{Kawato1999}
Kawato, M. (1999).
\newblock Internal models for motor control and trajectory planning.
\newblock {\em Current Opinion in Neurobiology}, 9(6):718--727.

\bibitem[Kikuchi, 1951]{Kikuchi1951}
Kikuchi, R. (1951).
\newblock A theory of cooperative phenomena.
\newblock {\em Physical Review}, 81(6):988--1003.

\bibitem[Koller, 2009]{Koller2009}
Koller, D. (2009).
\newblock {\em Probabilistic graphical models : principles and techniques}.
\newblock The MIT Press, Cambridge, Massachusetts.

\bibitem[Levine, 2018]{Levine2018}
Levine, S. (2018).
\newblock Reinforcement learning and control as probabilistic inference:
  Tutorial and review.
\newblock {\em arXiv:1805.00909}.

\bibitem[Lindig-León et~al., 2019]{LindigLeon2019}
Lindig-León, C., Gottwald, S., and Braun, D.~A. (2019).
\newblock Analyzing abstraction and hierarchical decision-making in absolute
  identification by information-theoretic bounded rationality.
\newblock {\em Frontiers in Neuroscience}, 13:1230.

\bibitem[Linson et~al., 2020]{Linson2020}
Linson, A., Parr, T., and Friston, K.~J. (2020).
\newblock Active inference, stressors, and psychological trauma: A
  neuroethological model of (mal)adaptive explore-exploit dynamics in
  ecological context.
\newblock {\em Behavioural Brain Research}, 380:112421.

\bibitem[Maccheroni et~al., 2006]{Maccheroni2006}
Maccheroni, F., Marinacci, M., and Rustichini, A. (2006).
\newblock Ambiguity aversion, robustness, and the variational representation of
  preferences.
\newblock {\em Econometrica}, 74(6):1447--1498.

\bibitem[MacKay, 2002]{MacKay2002}
MacKay, D. J.~C. (2002).
\newblock {\em Information Theory, Inference {\&} Learning Algorithms}.
\newblock Cambridge University Press, USA.

\bibitem[MacRae, 1970]{MacRae1970}
MacRae, A.~W. (1970).
\newblock Channel capacity in absolute judgment tasks: An artifact of
  information bias?
\newblock {\em Psychological Bulletin}, 73(2):112--121.

\bibitem[Marshall et~al., 2011]{Marshall2011}
Marshall, A.~W., Olkin, I., and Arnold, B.~C. (2011).
\newblock {\em Inequalities: Theory of Majorization and Its Applications}.
\newblock Springer New York, 2nd edition.

\bibitem[Mattsson and Weibull, 2002]{Mattsson2002}
Mattsson, L.-G. and Weibull, J.~W. (2002).
\newblock Probabilistic choice and procedurally bounded rationality.
\newblock {\em Games and Economic Behavior}, 41(1):61--78.

\bibitem[McFadden, 2005]{McFadden2005}
McFadden, D.~L. (2005).
\newblock Revealed stochastic preference: a synthesis.
\newblock {\em Economic Theory}, 26(2):245--264.

\bibitem[McKelvey and Palfrey, 1995]{McKelvey1995}
McKelvey, R.~D. and Palfrey, T.~R. (1995).
\newblock Quantal response equilibria for normal form games.
\newblock {\em Games and Economic Behavior}, 10(1):6--38.

\bibitem[Miller, 1956]{Miller1956}
Miller, G.~A. (1956).
\newblock The magical number seven, plus or minus two: some limits on our
  capacity for processing information.
\newblock {\em Psychological Review}, 63(2):81--97.

\bibitem[Minka, 2005]{Minka2005}
Minka, T. (2005).
\newblock Divergence measures and message passing.
\newblock Technical Report MSR-TR-2005-173, Microsoft.

\bibitem[Minka, 2001]{Minka2001}
Minka, T.~P. (2001).
\newblock Expectation propagation for approximate bayesian inference.
\newblock In {\em Proceedings of the 17th Conference in Uncertainty in
  Artificial Intelligence}, UAI '01, pages 362--369, San Francisco, CA, USA.
  Morgan Kaufmann Publishers Inc.

\bibitem[Mirza et~al., 2018]{Mirza2018}
Mirza, M.~B., Adams, R.~A., Mathys, C., and Friston, K.~J. (2018).
\newblock Human visual exploration reduces uncertainty about the sensed world.
\newblock {\em PLOS ONE}, 13(1):e0190429.

\bibitem[Mnih et~al., 2016]{Mnih2016}
Mnih, V., Badia, A.~P., Mirza, M., Graves, A., Lillicrap, T., Harley, T.,
  Silver, D., and Kavukcuoglu, K. (2016).
\newblock Asynchronous methods for deep reinforcement learning.
\newblock In Balcan, M.~F. and Weinberger, K.~Q., editors, {\em Proceedings of
  The 33rd International Conference on Machine Learning}, volume~48 of {\em
  Proceedings of Machine Learning Research}, pages 1928--1937, New York, New
  York, USA. PMLR.

\bibitem[Neal and Hinton, 1998]{Neal1998}
Neal, R.~M. and Hinton, G.~E. (1998).
\newblock A view of the em algorithm that justifies incremental, sparse, and
  other variants.
\newblock In Jordan, M.~I., editor, {\em Learning in Graphical Models}, pages
  355--368. Springer Netherlands, Dordrecht.

\bibitem[O'Donoghue et~al., 2020]{ODonoghue2020}
O'Donoghue, B., Osband, I., and Ionescu, C. (2020).
\newblock Making sense of reinforcement learning and probabilistic inference.
\newblock In {\em International Conference on Learning Representations}, ICLR
  ’20.

\bibitem[{Opper} and {Saad}, 2001]{Opper2001}
{Opper}, M. and {Saad}, D. (2001).
\newblock {\em Comparing the Mean Field Method and Belief Propagation for
  Approximate Inference in MRFs}, pages 229--239.

\bibitem[Ortega and Braun, 2013]{Ortega2013}
Ortega, P.~A. and Braun, D.~A. (2013).
\newblock Thermodynamics as a theory of decision-making with
  information-processing costs.
\newblock {\em Proceedings of the Royal Society A: Mathematical, Physical and
  Engineering Sciences}, 469(2153):20120683.

\bibitem[Ortega and Braun, 2014]{Ortega2014b}
Ortega, P.~A. and Braun, D.~A. (2014).
\newblock Generalized thompson sampling for sequential decision-making and
  causal inference.
\newblock {\em Complex Adaptive Systems Modeling}, 2(1):2.

\bibitem[Ortega and Stocker, 2016]{Ortega2016a}
Ortega, P.~A. and Stocker, A. (2016).
\newblock Human decision-making under limited time.
\newblock In {\em 30th Conference on Neural Information Processing Systems}.

\bibitem[Parr et~al., 2018]{Parr2018}
Parr, T., Benrimoh, D.~A., Vincent, P., and Friston, K.~J. (2018).
\newblock Precision and false perceptual inference.
\newblock {\em Frontiers in Integrative Neuroscience}, 12:39.

\bibitem[Parr and Friston, 2017]{Parr2017}
Parr, T. and Friston, K.~J. (2017).
\newblock Working memory, attention, and salience in active inference.
\newblock {\em Scientific reports}, 7(1):14678--14678.

\bibitem[Parr and Friston, 2019]{Friston2019}
Parr, T. and Friston, K.~J. (2019).
\newblock Generalised free energy and active inference.
\newblock {\em Biological Cybernetics}.

\bibitem[Parr et~al., 2019]{Parr2019}
Parr, T., Markovic, D., Kiebel, S.~J., and Friston, K.~J. (2019).
\newblock Neuronal message passing using mean-field, bethe, and marginal
  approximations.
\newblock {\em Scientific Reports}, 9(1):1889.

\bibitem[Pearl, 1988]{Pearl1988}
Pearl, J. (1988).
\newblock Belief updating by network propagation.
\newblock In Pearl, J., editor, {\em Probabilistic Reasoning in Intelligent
  Systems}, pages 143--237. Morgan Kaufmann, San Francisco (CA).

\bibitem[Poincaré, 1912]{Poincare1912}
Poincaré, H. (1912).
\newblock {\em Calcul des probabilités}.
\newblock Gauthier-Villars, Paris.

\bibitem[Powers, 1973]{Powers1973}
Powers, W.~T. (1973).
\newblock {\em Behavior: The Control of Perception}.
\newblock Aldine, Chicago, IL.

\bibitem[Rao and Ballard, 1999]{Rao1999}
Rao, R. P.~N. and Ballard, D.~H. (1999).
\newblock Predictive coding in the visual cortex: a functional interpretation
  of some extra-classical receptive-field effects.
\newblock {\em Nature Neuroscience}, 2(1):79--87.

\bibitem[Rosenkrantz, 1983]{Jaynes1983}
Rosenkrantz, R.~D. (1983).
\newblock {\em E.T. Jaynes: Papers on Probability, Statistics and Statistical
  Physics}.
\newblock Springer Netherlands, Dordrecht.

\bibitem[Russell and Subramanian, 1995]{Russel1995}
Russell, S.~J. and Subramanian, D. (1995).
\newblock Provably bounded-optimal agents.
\newblock {\em Journal of Artificial Intelligence Research}, 2(1):575--609.

\bibitem[Sales et~al., 2019]{Sales2019}
Sales, A.~C., Friston, K.~J., Jones, M.~W., Pickering, A.~E., and Moran, R.~J.
  (2019).
\newblock Locus coeruleus tracking of prediction errors optimises cognitive
  flexibility: An active inference model.
\newblock {\em PLOS Computational Biology}, 15(1):e1006267.

\bibitem[Saul and Jordan, 1996]{Saul1996}
Saul, L.~K. and Jordan, M.~I. (1996).
\newblock Exploiting tractable substructures in intractable networks.
\newblock In Touretzky, D.~S., Mozer, M.~C., and Hasselmo, M.~E., editors, {\em
  Advances in Neural Information Processing Systems 8}, pages 486--492. MIT
  Press.

\bibitem[Schach et~al., 2018]{Schach2018}
Schach, S., Gottwald, S., and Braun, D.~A. (2018).
\newblock Quantifying motor task performance by bounded rational decision
  theory.
\newblock {\em Frontiers in Neuroscience}, 12:932.

\bibitem[Schwartenbeck et~al., 2015]{Schwartenbeck2015}
Schwartenbeck, P., FitzGerald, T. H.~B., Mathys, C., Dolan, R., and Friston, K.
  (2015).
\newblock The dopaminergic midbrain encodes the expected certainty about
  desired outcomes.
\newblock {\em Cerebral cortex (New York, N.Y. : 1991)}, 25(10):3434--3445.

\bibitem[Schwartenbeck and Friston, 2016]{Schwartenbeck2016}
Schwartenbeck, P. and Friston, K. (2016).
\newblock Computational phenotyping in psychiatry: A worked example.
\newblock {\em eNeuro}, 3(4):ENEURO.0049--16.2016.

\bibitem[Schwöbel et~al., 2018]{Schwoebel2018}
Schwöbel, S., Kiebel, S., and Marković, D. (2018).
\newblock Active inference, belief propagation, and the bethe approximation.
\newblock {\em Neural Computation}, 30(9):2530--2567.

\bibitem[Shannon, 1948]{Shannon1948}
Shannon, C.~E. (1948).
\newblock A mathematical theory of communication.
\newblock {\em The Bell System Technical Journal}, 27:379--656.

\bibitem[Simon, 1955]{Simon1955}
Simon, H.~A. (1955).
\newblock A behavioral model of rational choice.
\newblock {\em The Quarterly Journal of Economics}, 69(1):99--118.

\bibitem[Sims, 2003]{Sims2003}
Sims, C.~A. (2003).
\newblock Implications of rational inattention.
\newblock {\em Journal of Monetary Economics}, 50(3):665--690.
\newblock Swiss National Bank/Study Center Gerzensee Conference on Monetary
  Policy under Incomplete Information.

\bibitem[Sims, 2016]{Sims2016}
Sims, C.~R. (2016).
\newblock Rate--distortion theory and human perception.
\newblock {\em Cognition}, 152:181--198.

\bibitem[Still, 2009]{Still2009}
Still, S. (2009).
\newblock Information-theoretic approach to interactive learning.
\newblock {\em Europhysics Letters}, 85(2):28005.

\bibitem[Tatikonda and Mitter, 2004]{TatikondaMitter2004}
Tatikonda, S. and Mitter, S. (2004).
\newblock Control under communication constraints.
\newblock {\em IEEE Transactions on Automatic Control}, 49(7):1056--1068.

\bibitem[Tenenbaum and Griffiths, 2001]{Tenenbaum2001}
Tenenbaum, J.~B. and Griffiths, T.~L. (2001).
\newblock Generalization, similarity, and bayesian inference.
\newblock {\em Behavioral and Brain Sciences}, 24(4):629--640.

\bibitem[Tinbergen, 1963]{Tinbergen1963}
Tinbergen, N. (1963).
\newblock On aims and methods of ethology.
\newblock {\em Zeitschrift für Tierpsychologie}, 20:410--433.

\bibitem[Tishby and Polani, 2011]{Tishby2011}
Tishby, N. and Polani, D. (2011).
\newblock Information theory of decisions and actions.
\newblock In Cutsuridis, V., Hussain, A., and Taylor, J.~G., editors, {\em
  Perception-Action Cycle: Models, Architectures, and Hardware}, pages
  601--636. Springer New York.

\bibitem[Todorov, 2008]{Todorov2008}
Todorov, E. (2008).
\newblock General duality between optimal control and estimation.
\newblock In {\em 2008 47th IEEE Conference on Decision and Control}. IEEE.

\bibitem[Todorov, 2009]{Todorov2009}
Todorov, E. (2009).
\newblock Efficient computation of optimal actions.
\newblock {\em Proceedings of the National Academy of Sciences},
  106(28):11478--11483.

\bibitem[Toussaint, 2009]{Toussaint2009}
Toussaint, M. (2009).
\newblock Robot trajectory optimization using approximate inference.
\newblock In {\em Proceedings of the 26th Annual International Conference on
  Machine Learning - {ICML} {\textquotesingle}09}. {ACM} Press.

\bibitem[Toussaint and Storkey, 2006]{Toussaint2006}
Toussaint, M. and Storkey, A. (2006).
\newblock Probabilistic inference for solving discrete and continuous state
  markov decision processes.
\newblock In {\em Proceedings of the 23rd International Conference on Machine
  Learning}, ICML ’06, pages 945--952, New York, NY, USA. Association for
  Computing Machinery.

\bibitem[von Neumann and Morgenstern, 1944]{Neumann1944}
von Neumann, J. and Morgenstern, O. (1944).
\newblock {\em Theory of Games and Economic Behavior}.
\newblock Princeton University Press, Princeton, NJ, USA.

\bibitem[Wainwright et~al., 2005]{Wainwright2005}
Wainwright, M., Jaakkola, T., and Willsky, A. (2005).
\newblock Map estimation via agreement on (hyper)trees: Message-passing and
  linear-programming approaches.
\newblock {\em IEEE Transactions on Information Theory}, 51(11):3697--3717.

\bibitem[Whittle, 1990]{Whittle1990}
Whittle, P. (1990).
\newblock {\em Risk-sensitive optimal control}.
\newblock Wiley, Chichester New York.

\bibitem[Wiener, 1948]{Wiener1948}
Wiener, N. (1948).
\newblock {\em Cybernetics: Or Control and Communication in the Animal and the
  Machine}.
\newblock John Wiley.

\bibitem[Williams, 1980]{Williams1980}
Williams, P.~M. (1980).
\newblock Bayesian conditionalisation and the principle of minimum information.
\newblock {\em The British Journal for the Philosophy of Science},
  31(2):131--144.

\bibitem[Williams and Peng, 1991]{Williams1991}
Williams, R.~J. and Peng, J. (1991).
\newblock Function optimization using connectionist reinforcement learning
  algorithms.
\newblock {\em Connection Science}, 3(3):241--268.

\bibitem[Winn and Bishop, 2005]{Winn2005}
Winn, J. and Bishop, C.~M. (2005).
\newblock Variational message passing.
\newblock {\em J. Mach. Learn. Res.}, 6:661--694.

\bibitem[Wolpert, 2006]{Wolpert2006}
Wolpert, D.~H. (2006).
\newblock {\em Information Theory -- The Bridge Connecting Bounded Rational
  Game Theory and Statistical Physics}, pages 262--290.
\newblock Springer Berlin Heidelberg.

\bibitem[Wolpert, 2019]{Wolpert2019}
Wolpert, D.~H. (2019).
\newblock The stochastic thermodynamics of computation.
\newblock {\em Journal of Physics A: Mathematical and Theoretical},
  52(19):193001.

\bibitem[Yedidia et~al., 2001]{Yedidia2001}
Yedidia, J.~S., Freeman, W.~T., and Weiss, Y. (2001).
\newblock Generalized belief propagation.
\newblock In Leen, T.~K., Dietterich, T.~G., and Tresp, V., editors, {\em
  Advances in Neural Information Processing Systems 13}, pages 689--695. MIT
  Press.

\bibitem[Yedidia et~al., 2005]{Yedidia2005}
Yedidia, J.~S., Freeman, W.~T., and Weiss, Y. (2005).
\newblock Constructing free-energy approximations and generalized belief
  propagation algorithms.
\newblock {\em IEEE Transactions on Information Theory}, 51(7):2282--2312.

\bibitem[Yuille and Kersten, 2006]{Yuille2006}
Yuille, A. and Kersten, D. (2006).
\newblock Vision as bayesian inference: analysis by synthesis?
\newblock {\em Trends in Cognitive Sciences}, 10(7):301--308.
\newblock Special issue: Probabilistic models of cognition.

\bibitem[Yuille, 2002]{Yuille2002}
Yuille, A.~L. (2002).
\newblock Cccp algorithms to minimize the bethe and kikuchi free energies:
  Convergent alternatives to belief propagation.
\newblock {\em Neural Computation}, 14(7):1691--1722.

\bibitem[Ziebart, 2010]{Ziebart2010}
Ziebart, B.~D. (2010).
\newblock {\em Modeling Purposeful Adaptive Behavior with the Principle of
  Maximum Causal Entropy}.
\newblock PhD thesis, Carnegie Mellon Unversity.

\end{thebibliography}

}

\appendix
\section{Appendices}

\subsection{Derivation of exemplary update equations}
\label{S1_App}

\subsubsection{$Q$-value Active Inference}
In the simple example of Section 5.2 under the partial mean-field assumption (23), and in the case when the desired distribution $p_\mathrm{des}$ is combined with the generative model $p_0$ via the value function $Q$ as shown in Equation (24), i.e.~if $\phi\propto p_0(x,X',\mathbf{S},A) \, e^{Q(A)}$, then the full free energy $F(q\|\phi)$ can be written as
\begin{align}  \nonumber
& F(q\|\phi)  = F(q(\mathbf{S}|A)q(A)\|p_0(x|S) p_0(\mathbf{S}|A) p_0(A) e^{Q(A)})\\
    & = \big \langle F_\mathbf{S}(A) - Q(A)\big \rangle_{q(A)} + D_{\mathrm{KL}}(q(A)\|p_0(A)) \label{S1} 
\end{align}
where, $F_\mathbf{S}(A) - Q(A)$ is given by
\begin{equation}  \label{FminusQ}
\left \langle \log \frac{~ \qquad q(S) \quad q(S'|A) \quad  \sum_{s'}p_0(X'|s')q(s'|A)}{p_0(x|S)p_0(S) \, p_0(S'|S,A) \, p_\mathrm{des}(X') p_0(X'|S')} \right \rangle
\end{equation}
where the expectation is with respect to $q(X',\mathbf{S}|A)$. Thus, optimizing \eqref{S1} over $q(A)$, while keeping $q(\mathbf{S}|A)$ fixed, results in a Boltzmann distribution with prior $p_0(A)$ and energy $F_\mathbf{S}(A) - Q(A)$. When optimizing $F(q\|\phi)$ with respect to $q(S)$ while keeping $q(S'|A)$ and $q(A)$ fixed, we have 
\begin{align} \nonumber
& q^\ast(S)  = \argmax_{q(S)} \, F(q\|\phi) = \argmax_{q(S)} \,\langle F_\mathbf{S}(A)\rangle_{q(A)} \\
          & = \argmax_{q(S)} \, \underbrace{\left\langle \log \frac{q(S)}{p_0(x|S)p_0(S) e^{\langle T \rangle_{q(S'|A)q(A)}}} \right\rangle_{q(S)}}_{F\big(q(S)\big\|p_0(x|S)p_0(S) e^{\langle T \rangle}\big)} \, , \label{S2}
\end{align}
where $T\,{\coloneqq}\, \log p_0(S'|S,A)$ is shorthand for the log-transition probability. Hence, from \eqref{S2} we can read off the solution $q^\ast(S)$ in virtue of the general optimum (14) of variational free energy. While here it was enough to optimize $\langle F_\mathbf{S}\rangle_q$, because in contains the only dependencies of $F(q\|\phi)$ on $q(S)$, this is not the case for $q(S'|A)$, since also $Q$ depends on $q(S'|A)$. Thus, when optimizing \eqref{S1} over $q(S'|A)$ while keeping $q(A)$ and $q(S)$ fixed, one has to optimize $\langle F_\mathbf{S}-Q\rangle$ which does not take the form of a free energy in $q(S'|A)$ due to the functional dependency of $q(X'|A) = \sum_{s'}p_0(X'|s')q(s'|A)$ on $q(S'|A)$ that appears in \eqref{FminusQ}. However, this type of dependency is largely ignored in the Active Inference literature (as for example noted in the appendix of \citep{Friston2015}), since the optimization with respect to $q(S'|A)$ would not have a closed-form solution otherwise. 


Once this term is ignored, then the objective for $q(S'|A)$ takes a very simple form,
\begin{align} 
q^\ast(S'|A) & = \argmax_{q(S'|A)} \, F(q\|\phi) \nonumber \\ \label{S3}
& \approx \argmax_{q(S'|A)} \left \langle \log \frac{q(S'|A)}{e^{\langle T\rangle_{q(S)}}}\right\rangle_{q(S'|A)} ,
\end{align}
from which we can again read off the resulting update equation. In total, from \eqref{S1},\eqref{S2}, and \eqref{S3} we obtain the set of equations
\begin{subequations} \label{eq:updateequations}
\begin{align} \label{qStarQvalue}
q^\ast(S) & = \tfrac{1}{\mathcal{Z}}\, p_0(x|S) p_0(S) e^{\langle T\rangle_{q(S'|A)q(A)}} \\ 
q^\ast(S'|A) & \approx \tfrac{1}{\mathcal{Z}(A)}\, e^{\langle T\rangle_{q(S)}}\\ 
q^\ast(A) &= \tfrac{1}{\mathcal{Z}} \, p_0(A) e^{-F_\mathbf{S}(A)+Q(A)},
\end{align}
\end{subequations}
where $\mathcal{Z}$ denotes the respective normalization constants and $T\eqq \log p_0(S'|S,A)$.

It is important to note, however, that update equations in Active Inference resulting from a mean-field assumption (even if it is a partial mean-field assumption such as (23)) should  be taken with care, since---as is demonstrated in the grid world simulations in S2 Notebook---even in very simple situations the resulting agents fail to correctly plan actions that lead to desired states.

\medskip

\subsubsection{Direct Active Inference (variational Control as Inference)---mean-field assumption}
Here, we derive the update equations resulting from the minimization of the variational free energy for the reference defined in Equation (25a), i.e.~a variational formulation of Control as inference \citep{Toussaint2006}, under the mean-field assumption (23). We start by writing the variational free energy $F(q\|\phi)$ in a form analogous to \eqref{S1}, where now $\phi$ is given by $p_0(X'|S')p_0(x,\mathbf{S}|A)p_0(A)$,  
\begin{align*}
F(q\|\phi)& = \Big\langle \underbrace{F(q(\mathbf{S}|A)\|p_0(x,\mathbf{S}|A))}_{=F_\mathbf{S}(A)} - G(A) \Big\rangle_{q(A)} \\
& \quad  + D_\mathrm{KL}(q(A)\|p_0(A)) \, ,
\end{align*}
where 
\[
G(A) \coloneqq \Big\langle \underbrace{\langle \log p_\mathrm{des}(X')\rangle_{p_0(X'|S')}}_{\eqqcolon g(S')}\Big\rangle_{q(S'|A)}\,.
\]
Note that, compared to $Q$-value Active Inference, here we do not have to make any additional approximations, because $G$ only depends linearly on $q(S'|A)$. 

Similarly to above, when optimizing with respect to $q(A)$ while keeping $q(S)$ and $q(S'|A)$ fixed, we obtain that $q^\ast(A)$ is a Boltzmann distribution with energy $F_\mathbf{S}-G$ and prior $p_0(A)$. Optimizing $q(S)$ while keeping $q(A)$ and $q(S'|A)$ constant has the same result as shown in \eqref{qStarQvalue} because as before the only dependencies on $q(S)$ are in $F_\mathbf{S}$. Finally, in order to read off the solution of the optimization with respect to $q(S'|A)$ while keeping $q(S)$ and $q(A)$ constant, we can rewrite $F_\mathbf{S}-G$ as follows
\begin{align*}
q^\ast(S'|A) & = \argmax_{q(S'|A)}\, F(q\|\phi) = \argmax_{q(S'|A)} \big(F_\mathbf{S}(A)-G(A)\big) \\
& =  \argmax_{q(S'|A)} \left\langle  \log \frac{q(S'|A)}{e^{\langle T \rangle_{q(S)}+g(S')}}  \right\rangle_{q(S'|A)} 
\end{align*}
so that in total we obtain the set of equations
\begin{subequations} \label{eq:updateequationsGF} \
\begin{align} 
q^\ast(S) & = \tfrac{1}{\mathcal{Z}}\, p_0(x|S) p_0(S) e^{\langle T\rangle_{q(S'|A)q(A)}} \\ 
q^\ast(S'|A) & = \tfrac{1}{\mathcal{Z}(A)}\, e^{\langle T\rangle_{q(S)}+g(S')}\\ 
q^\ast(A) &= \tfrac{1}{\mathcal{Z}} \, p_0(A) e^{-F_\mathbf{S}(A)+G(A)},
\end{align}
\end{subequations}
where $\mathcal{Z}$ denotes the respective normalization constants, and again $T=\log p_0(S'|S,A)$.

It is noteworthy that recently another free energy approach similar to Active Inference has been introduced that does not make use of variational free energy, but of a different functional termed \textit{generalized free energy}  \citep{Friston2019}. Despite of the different functional form, this version uses a reference function that is similar to the direct Active Inference approach, where the desired distribution is also multiplied directy to the generative model but with a renormalization that results in a modified generative model over observations, states, and actions. Using this renormalized reference in a variational free energy approach would result in trivial inference reproducing the fixed prior $p_0(A)$, corresponding to Bayes' conditioning the modified generative model on the past analogous to perceptual Bayesian inference, e.g., $p(A|X) = p_0(A)$ in the case of the one-step example. In contrast, the minimization of the free energy functional used in \citep{Friston2019} does not correspond to a Bayesian inference process, which is why we do not further discuss it here.

\subsubsection{Direct Active Inference (variational Control as Inference)---Bethe assumption}

Here, we derive the update equations resulting from the minimization of the variational free energy for the reference (25a) under a Bethe approximation, which therefore is a more precise variational formulation of Control as Inference as the mean-field approximation of the previous section. In fact, it turns out that such equations are equivalent to Belief propagation \citep{Yedidia2001}, a well-known inference method that produces exact marginals in tree-like graphs \citep{Pearl1988}, such as the probabilistic models considered in the article and in the Active Inference literature.

Analogous to the previous section, without any specific restrictions on $q$ we can write the total free energy for the one-step example from Section 5.2 with the reference (25a) as 
\begin{align*}
F(q\|\phi) & = \underbrace{\left \langle \log \frac{q(X',S,S'|A)}{p_0(R\eqq 1,X \eqq x,X',S,S'|A)} \right\rangle_{q}}_{\eqqcolon \, \langle F(A)\rangle_{q(A)}} \\ 
& \quad + \, D_\mathrm{KL}(q(A)\|p_0(A))
\end{align*}
from which it immediately follows that minimizing with respect to $q(A)$, while considering $q(X',S,S'|A)$ constant, results in a Boltzmann distribution with energy $F(A)$ and prior $p_0(A)$. $F(A)$ is the variational free energy of $q(X',S,S'|A)$ with respect to the reference $p_0(R\eqq 1,X \eqq x,X',S,S'|A)$ given by 
\[
\underbrace{p_0(x|S) p_0(S)}_{\eqqcolon f_1(S)} \underbrace{p_0(S'|S,A)}_{\eqqcolon f_2(S,S')} \underbrace{p_0(X'|S')}_{\eqqcolon f_3(S',X')} \underbrace{p_\mathrm{des}(X')}_{\eqqcolon f_4(X')} \,.
\]
Thus, minimizing $F(A)$ with respect to $q(X',S,S'|A)$ without any restrictions or simplifications results in the exact Bayes' posterior $p(X',S,S'|A,R=1,X=x)$ 
\[
\frac{1}{\mathcal{Z}(A)} \, f_1(S) \, f_2(S,S') \, f_3(S',X') \, f_4(X') \, ,
\]
where $\mathcal{Z}(A)$ denotes the corresponding normalization< constant. The problem that we want to solve is to find an approximation to this Bayes' posterior that is more precise than the mean-field approximation of the previous section but requires less involved computations than the determination of $\mathcal{Z}(A)$. While one attempt is to partition the full graph into smaller graphs and apply a naive mean-field approximation inside of each subgraph, known as a \textit{structured} mean-field approximation \citep{Saul1996}, the Bethe approximation follows a slightly different approach. It is the simplest version of the \textit{cluster variation methods} often attributed to Kikuchi \citep{Kikuchi1951}, a family of region-based free energy approximations \citep{Yedidia2005}, where one keeps beliefs over different sections of the factor graph. Specifically, in the Bethe assumption, the regions consist of each factor and its neighbouring nodes, which can also be seen as allowing pair-wise interactions. Following the systematic treatment in \citep{Yedidia2005}, the Bethe approximation for our example consists of seven belief functions, one for each factor, $b_1, \dots, b_4$, and one for each variable, $b_S$, $b_{S'}$, and $b_{X'}$,
\begin{equation} \label{eq:Betheapprox}
q(S,S',X'|A) = \frac{b_1(S) b_2(S,S') b_3(S',X') b_4(X')}{b_S(S) b_{S'}(S') b_{X'}(X')}
\end{equation}
where the marginals of the factor beliefs are required to be consistent with the single-variable beliefs. Thus, the variational free energy $F(A)$ can be written as
\[
F(A) = \sum_{k=1}^4 \left\langle \log \frac{b_k}{f_k} \right\rangle_{b_k} - \sum_{Y\in\{S,S',X'\}} \langle \log b_Y \rangle_{b_Y}
\]
which has to be minimized under the consistency and normalization contraints, leading to the Lagrangian 
\begin{align*}
F(A)& + \sum_s\lambda_1(s)\big(b_S(s)-b_1(s)\big)  \\
& + \sum_s \lambda_{2S}(s)\left(b_S(s)-\sum_{s'}b_2(s,s') \right) \\
& + \sum_{s'} \lambda_{2S'}(s')\left(b_{S'}(s')-\sum_{s}b_2(s,s') \right)\\
& + \sum_{s'} \lambda_{3S'}(s')\left(b_{S'}(s')-\sum_{x'}b_3(s',x') \right) \\
& + \sum_{x'} \lambda_{3X'}(x')\left(b_{X'}(x')-\sum_{s'}b_3(s',x') \right)\\
& + \sum_{x'}\lambda_4(x')\big(b_{X'}(x')-b_4(x')\big)  \\
& + \sum_{k=1}^4 \gamma_k \left(\sum b_k - 1 \right) + \sum_{Y\in\{S,S',X'\}} \gamma_Y \left( \sum b_Y - 1 \right)
\end{align*}
where the Lagrange multipliers for the consistency constraints are denoted by $\lambda$ and the Lagrange multipliers for the normalization constraints by $\gamma$. The equations for the beliefs at the stationary points (zeroes of the derivatives of the Lagrangian) are 
\begin{align*}
b_1(s) & \ \propto \ f_1(s) \, e^{\lambda_1(s)} \,, \\
b_2(s,s') & \ \propto \ f_2(s,s') \, e^{\lambda_{2S}(s)} \, e^{\lambda_{2S'}(s')}\, , \\
b_3(s',x') & \ \propto \ f_3(s',x') \, e^{\lambda_{3S'}(s')} \, e^{\lambda_{3X'}(x')} \, , \\
b_4(x') & \ \propto \ f_4(x') \, e^{\lambda_4(x')} \, ,\\
b_S(s) &  \ \propto \ e^{\lambda_1(s)} \, e^{\lambda_{2S}(s)}\, , \\
b_{S'}(s') & \ \propto \ e^{\lambda_{2S'}(s')} \, e^{\lambda_{3S'}(s')}\, , \\
b_{X'}(x') & \ \propto \ e^{\lambda_{3X'}(x')} \, e^{\lambda_4(x')} \, , 
\end{align*}
where the proportionality sign $\propto$ means that the left-hand side results from normalizing the right hand-side to obtain a probability distribution. By writing $m_l \coloneqq e^{\lambda_l}$ for all $l\in\{1,2S,2S',3S',3X',4\}$, we obtain from the stationarity conditions and the consistency constraints
\begin{subequations} \label{eq:messages}
\begin{align}
m_{2S}(s) \ & \propto \ f_1(s)\\
m_{1}(s) \ & \propto \ \sum\nolimits_{s'} f_2(s,s') \,  m_{2S'}(s') \\
m_{3S'}(s') \ & \propto \ \sum\nolimits_{s} f_2(s,s') \, m_{2S}(s) \\
m_{2S'}(s') \ & \propto \ \sum\nolimits_{x'} f_3(s',x') \, m_{3X'}(x') \\
m_4(x') \ & \propto \ \sum\nolimits_{s'} f_3(s',x') \, m_{3S'}(s') \\
m_{3X'}(x') \ & \propto \ f_4(x')\, .
\end{align}
\end{subequations}
The update equations for the beliefs in \eqref{eq:Betheapprox} can be obtained by iterating the equations in \eqref{eq:messages} and using the stationarity conditions that express the beliefs in terms of the $m_l$. Note that the quantities denoted by $m_l$ are usually interpreted as local messages that are sent between the nodes and factors of the underlying graphical model \citep{Yedidia2005}, e.g., $m_{3S'}$ is considered a message sent from node $S'$ to factor $3$, which can be used to determine the message $m_4$ from factor $3$ to node $X'$ by weighing with $f_3$ and summing over $S'$, etc. By this identification, variational inference under the Bethe approximation is equivalent to belief propagation. While in \eqref{eq:messages} there is at most one message that is multiplied to the factor $f_k$ before the sum is taken, in more complex factor graphs, where more than 2 nodes are connected to a factor, the messages coming in to a factor from the neighboring nodes are multiplied before they are summed to calculate the outgoing message, which is why this type of message-passing is also known as the \textit{sum-product} algorithm.

\subsection{Details on the example in Fig 8}
\label{S2_App}

Here, we are giving additional details on Figure 8 in the article. We consider the simple example of three possible observations, $x_1,x_2,x_3$, a desired distribution $p_\mathrm{des}(X') = (1/3,1/6,1/2)$, two actions $a$ with predictive distributions $p(X'|A\eqq 1) \eeqq (1,0,0)$ and $p(X'|A\eqq 2)\eeqq (0,1/2,1/2)$, and a constant prior $p_0(A) = (1/2,1/2)$. We can consider $p(X'|A)$ as a result of marginalizing the generative model $p_0(X',S',A) = p_0(X'|S')p_0(S'|A)p_0(A)$ with state distributions $p(S'|A\eqq 1) \eeqq (1,0,0)$ and $p(S'|A\eqq 2)\eeqq (0,1/2,1/2)$ and an emission probability $p_0(X'|S')$ that is chosen such that the given $p(X'|A)$ equals $p(X'|A) = \sum_{s'} p_0(X'|s') p_0(s'|A)$. Suitable emission probabilities have for example the form $p_0(X'\eqq x_i|S' \eqq s_j) = M_{ij}(t)$, where
\[
M(t) = \left( \begin{array}{ccc} 1 & 0 & 0 \\ 0 & t & 1{-}t \\ 0 & 1{-}t & t \end{array}\right)
\]
for all $t\in[0,1]$. Note that the resulting average entropies $\langle H(p_0(S'|A\eqq2))\rangle_{p_0(S'|A=2)}$ are in the range $[0,1]$ bit for $A=2$ (always zero for $A=1$), where the extreme values are assumed at $t\in\{0,1\}$ (0 bit) and $t=1/2$ (1 bit).
 
Furthermore, for the application of Active Inference in Figure 8, we have considered an exact version of the value function, $Q\eqq Q_\mathrm{exact}$, where the trial distribution $q(S'|A)$ is replaced by the exact predictive distribution $p_0(S'|A)$. In this ``exact'' interpretation, the corresponding action distributions $p(A)\propto p_0(A) e^{Q_\mathrm{exact}(A)}$ could then be viewed as defining the ideal behaviour that is approximated by the variational free energy minimization. In the Active Inference literature, $p(A)\propto p_0(A) e^{Q(A)}$ is considered a ``prior'', because it is viewed as part of the generative model and thus is part the input to the variational inference process. However, by considering $Q$ an approximation of $Q_\mathrm{exact}$ these distributions can be viewed as defining the ideal behavior that is approximated by the trial distributions during free energy minimization and are therefore more in line with the ``posteriors'' in other decision-making models (even though the value function $Q(A)$---and therefore $p(A)$---is presupposed, in constrast to being the result of some principle).

\subsection{Surprise minimization}
\label{S3_App}

The (\textit{informational}) \textit{surprise} or \textit{surprisal} of a given element $x$ with respect to a probability distribution $p_0(X)$ is defined as $S_0\coloneqq {-}\hspace{-1pt}\log p_0(x)$, i.e.~it is simply a strictly decreasing function of probability such that outcomes $x$ with low probability have high surprise and outcomes $x$ with high probability have low surprise. A common statement found in the literature \citep{Parr2017} is that variational free energy is an upper bound on surprise and thus minimizing free energy also minimizes surprise. This idea originates from the special case of greedy inference with latent variables, where, for fixed data $x$, the goal is to maximize the likelihood $p_\theta(x)=\sum_z p_\theta(x,z)$ with respect to a parameter $\theta$. If the marginalization over the latent variable $Z$ is too hard to carry out directly, then one might take advantage of the bound
\begin{equation} \label{eq:boundOnSurpriseEM}
F(q(Z)\|p_\theta(x,Z)) \geq -\log p_\theta(x) \eqqcolon S_\theta ,
\end{equation}
i.e.~that the variational free energy of $q(Z)$ is an upper bound on the surprise $S_\theta$, which might therefore be reduced by minimizing its upper bound \textit{with respect to $\theta$} as a proxy. In the variational Bayes' approach to the above inference problem, where $\theta$ is treated as a random variable $\Theta$, minimization with respect to $\theta$ is replaced by the minimization with respect to $q(\Theta)$. In this case, the analogous bound to \eqref{eq:boundOnSurpriseEM} is
\[
F(q(Z|\Theta)q(\Theta)\|p_0(x,Z,\Theta)) \geq -\log \sum_z e^{\langle \log p_0(x,z,\Theta)\rangle_{q(\Theta)}} ,
\]
where the right-hand side is the minimum of the left-hand side with respect to $q(Z|\Theta)$. In this sense, variational free energy is generally not a bound on the surprise $S_\Theta$ anymore, but on a log-sum-exp version of it instead. Nonetheless, also in this Bayesian approach, variational free energy is an upper bound on the surprise $S_0$,
\begin{equation} \label{eq:boundOnSurprise}
F(q(Z|\theta)q(\theta)\|p_0(x,Z,\Theta)) \geq - \log p_0(x) = S_0,
\end{equation}
where the right-hand side is the minimum of the left-hand side with respect to both $q(Z|\theta)$ and $q(\theta)$. However, in contrast to \eqref{eq:boundOnSurpriseEM}, there is no variable left in $S_0$ over which one could minimize. Therefore, saying that minimizing free energy also minimizes surprise \citep{Parr2017}, is generally only true in the sense that minimizing free energy minimizes an upper bound on surprise, however surprise itself is not minimized. Instead, the important fact about \eqref{eq:boundOnSurprise} is that equality is achieved by the Bayes' posteriors $q(Z|\Theta)=p_0(Z|\Theta,x)$ and $q(\Theta)=p_0(\Theta|x)$ as discussed in Section 3.2.1.

\subsection{Separation of model and state variables}
\label{S4_App}

In $Q$-value Active Inference, action and perception do not optimize the same variational free energy but two different free energy expressions. This is motivated from the separation of model variables $M$ and state variables in standard variational Bayesian inference, where the full free energy can be split up into a sum of a state free energy $F_M$ averaged over models $M$ and a KL term that is independent of state distributions. Optimizing the full free energy can then be done separately by alternatingly doing perceptual inference by optimizing $F_M$ for each model $M$ and optimizing the full free energy to find the model distribution $q(M)$. In Active Inference, where actions $A$ might be thought of analogous to models $M$ in Bayesian inference, the full free energy is analogously split up into a sum of a state free energy $F_A$ averaged over actions $A$ and a KL term which---in contrast to standard Bayesian inference---\textit{does} depend on state distributions. However, Active Inference essentially ignores this extra $q$-dependency by following the analogous optimization scheme to Bayesian inference: one alternatingly optimizes $F_A$ with respect to state distributions and then the full free energy with respect to the action distributions $q(A)$. In particular, this separation into state and action free energies is not a consequence of optimizing the full variational free energy, but a deliberate choice made by Active Inference. 

In the following, we discuss in more detail how this separation follows from optimizing the full free energy in standard Bayesian inference and highlight how $Q$-value Active Inference adopts the same optimization scheme but by giving up the optimization of a single variational free energy.

\subsubsection{Bayesian inference}

Consider the case of multiple probabilistic models $p_m(X,Z)$ that are indexed by a label $m$, where each $p_m$ describes a different probabilistic relationship between data $X$ and hidden states $Z$. Given data $X=x$, one could find the best $m$ by selecting the model with the largest marginal likelihood $p_m(x) = \sum_z p_m(x,z)$. A popular method to accomplish this is the basic EM algorithm \cite{Dempster1977}, where $m$ is optimized greedily while $Z$ is inferred using Bayesian inference for a given $m$ (either exact or approximate). In a purely Bayesian treatment, one also assumes a prior distribution over models $p_0(M)$, so that the full joint over data $X$, hidden states $Z$, and models $M$ becomes
\[
p_0(X,Z,M) \coloneqq  p_M(X,Z)  p_0(M) \eqqcolon  p_0(X,Z|M)\, p_0(M) \, . 
\]
The Bayes' posterior $p(M|X)$ can then simply be determined from the Bayes' posterior $p(Z,M|X)$ through marginalization over $Z$. As discussed in the article (Section 3.2), if direct Bayesian inference is infeasable then a variational formulation might be useful, where trial distributions $q(Z,M)$ over the unknown variables $M$ and $Z$ are fitted to the reference $\phi(Z,M) \coloneqq p_0(x,Z,M)$ by minimizing the variational free energy 
\[
F(q\|\phi) = \left \langle \log \frac{q(Z,M)}{p_0(x,Z,M)} \right\rangle_{q(Z,M)} \, .
\]
By writing $q$ and $p_0$ in their factorized forms
\begin{align*}
q(Z,M) & = q(Z|M) q(M), \\
p_0(X,Z,M) & = p_0(X,Z|M) p_0(M) \, ,
\end{align*}
the variational free energy can be decomposed as 
\begin{align} \nonumber
F(q\|\phi) & = \left\langle \log \frac{q(Z|M) q(M) }{p_0(x,Z|M) p_0(M)} \right \rangle_{q} \\ \nonumber
& = \Big\langle \underbrace{\Big\langle \log \frac{q(Z|M)}{p_0(x,Z|M)} \Big\rangle_{q(Z|M)}}_{\eqqcolon F_M} \Big\rangle_{q(M)} \\ 
& \quad  + \Big\langle \log \frac{q(M)}{p_0(M)}\Big\rangle_{q(M)} \nonumber \\[5pt] \label{splitup}
& = \langle F_M \rangle_{q(M)} + D_\mathrm{KL}(q(M)\|p_0(M))  \, .
\end{align}
Notably, the minimization of $F$ with respect to $q$ splits up into the minimization of the free energy over states
\begin{equation} \label{freeenergyoverstates}
F_M = F\big(q(Z|M) \big\|p_0(x,Z|M)\big)  
\end{equation}
with respect to $q(Z|M)$, and the minimization of \eqref{splitup} with respect to $q(M)$. 
In particular, the inference over models and states, $(M,Z)$, separates into inference over hidden states for each model, which determines $F_M$ for each $M$, and inference over $M$.

\medskip
\subsubsection{Active Inference}
In $Q$-value Active Inference, action selection is treated similarly to model selection in Bayesian inference discussed in the previous section. However, the KL term in \eqref{splitup} also depends on trial distributions over states, which means that a separation into action and state variables analogous to the separation in model selection is not possible when considering the problem of action and perception as the minimization of a single free energy functional, which is usually the conceptual starting point in the Active Inference literature \cite{Friston2010,Friston2018}.

More precisely, as discussed in Section 5, the reference function $\phi$ that enters the variational free energy in $Q$-value Active Inference is constructed from a given probabilistic model $p_0$ and a value function $Q$ by replacing the fixed prior $p_0(A)$ over actions with the modified distribution $\tilde p_0(A)\coloneqq \frac{1}{\mathcal{Z}}\, p_0(A) e^{Q(A)}$. As can be seen exemplarily in the one-step case in Equation (26), the value function $Q$ depends on trial distributions $q$ over hidden states and therefore $\tilde p_0(A)$ depends on $q$ as well. Despite this dependency, the total free energy $F(q\|\phi)$ can still be written as 
\begin{align} 
F(q\|\phi) & = \left\langle \log \frac{q(X',\mathbf{S}|A) q(A)}{p_0(x,X',\mathbf{S}|A) \tilde p_0(A)}\right \rangle_{q} \nonumber \\ \label{splitup-actinf}
& = \langle F_A \rangle_{q(A)} + D_\mathrm{KL}\big(q(A)\|\tilde p_0(A)\big)
\end{align}
with
\begin{equation} \label{freeenergyoverstates-actinf}
F_A \coloneqq F\big(q(X',\mathbf{S}|A)\big\| p_0(x,X',\mathbf{S}|A) \big) \, .
\end{equation}
in the case of the one-step example of Section 5.2. Equations \eqref{splitup-actinf} and \eqref{freeenergyoverstates-actinf} are analogous to Equations \eqref{splitup} and \eqref{freeenergyoverstates}, respectively. However, when optimizing the full free energy $F(q\|\phi)$ with respect to the factor $q(X',\mathbf{S}|A)$, one would have to consider both terms in the decomposition \eqref{splitup-actinf}, since, unlike $p_0(M)$ in the previous section, here $\tilde p_0(A)$ does depend on trial distributions over states (the factor $q(S'|A)$). It should be noted that this dependency is non-linear and non-local, and therefore a closed-form solution cannot be derived (cf. $(ii)$ in Section 5.3). 

In Active Inference, this complication is avoided by simply ignoring the $q$-dependency of $Q$ when deriving the update equations, or, put differently, one optimizes two different free energies for perception and action: one first optimizes $F_A$ with respect to state distributions for each action $A$ and then one optimizes the full free energy \eqref{splitup-actinf} with respect to $q(A)$. This is in analogy to Bayesian model selection of the previous section, where this separation was a consequence of the minimization of the full free energy. However, here, due to the extra dependency of $\tilde p_0(A)$ on $q$, it is not a consequence but a choice made by Active Inference. This means that one no longer does variational inference over the combined set of states and actions, but variational inference over states with free energy $F_A$ and variational inference over actions with free energy \eqref{splitup-actinf}. In particular, there is not a single free energy that is optimized by both perception and action, but two different ones.

\enlargethispage{1cm}
\setcounter{section}{18}

\section{Supplementary Material}

The following ancillary files are provided as supplementary material:


\subsection{Notebook: Comparison of different formulations of Active Inference}
\label{S1_Comp} A detailed comparison of the different formulations of Active Inference found in the literature (2013-2018), including their mean-field and exact solutions in the general case of arbitrary many time steps. 

\subsection{Notebook: Grid world simulations}
\label{S1_Sim} 
We provide implementations of the models discussed in this article in a grid world environment, both as a rendered html file as well as a jupyter notebook that is \href{https://github.com/sgttwld/two-kinds-of-free-energy--notebooks}{available on github}.

\end{document}